\usepackage{graphicx, url, amsmath, setspace, multirow}
\pdfoutput=1
\documentclass{aa}
\newcommand{\Mach}{$\mathcal{M}$}
\newcommand{\Tex}{$T_{\rm ex}$}
\newcommand{\Tkin}{$T_{\rm kin}$}
\newcommand{\vturb}{$v_{\rm turb}$}

\newcommand{\vel}[1]{#1~m\,s$^{-1}$}

\newcommand{\new}[1]{{\bf #1}\,}

\newcommand{\FigureCNCSMaps}{
\begin{figure*}
\centering
\includegraphics[width=1.9\columnwidth]{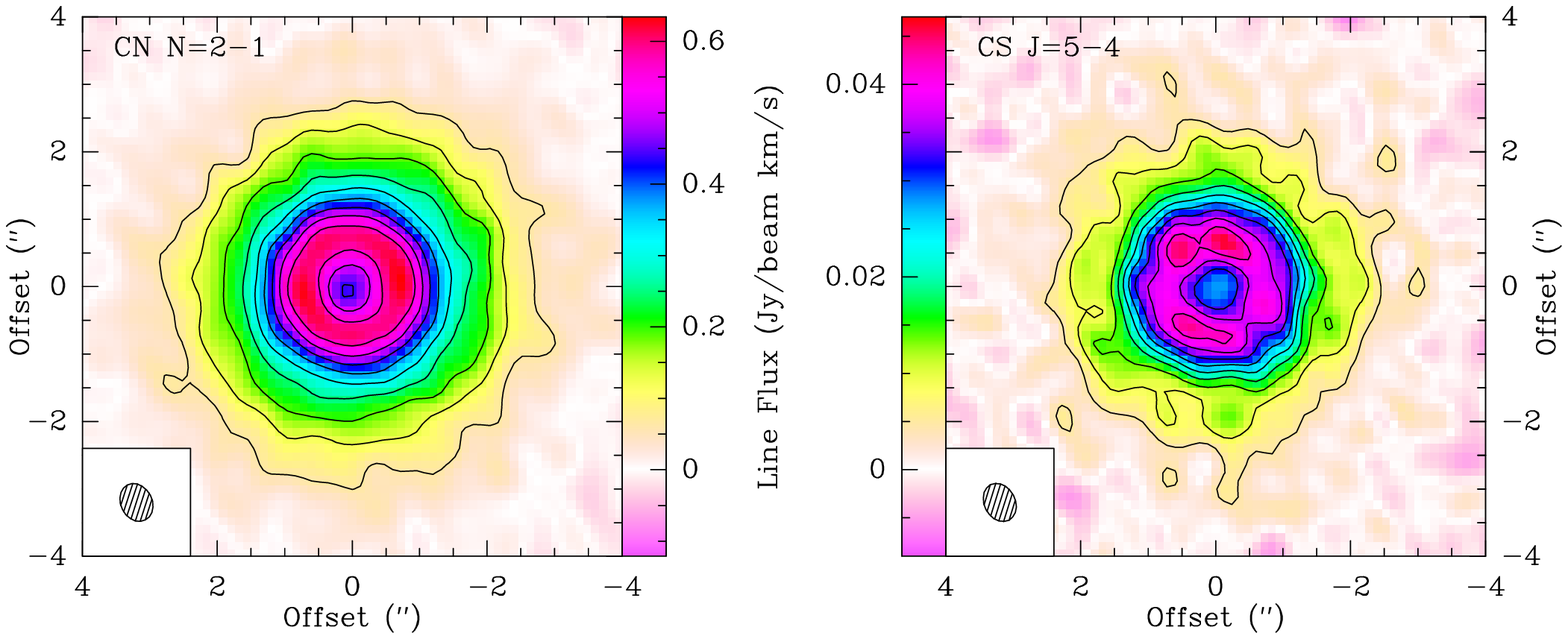}
\caption{Integrated intensity maps of CN including all hyperfine components, left, and CS, right. Contours are 10~\% of the peak value. No azimuthal structure is seen within the noise for both lines.}
\label{fig:CNCS_integratedintensity}
\end{figure*}
}

\newcommand{\FigureTemperatureBias}{
\begin{figure}
\includegraphics{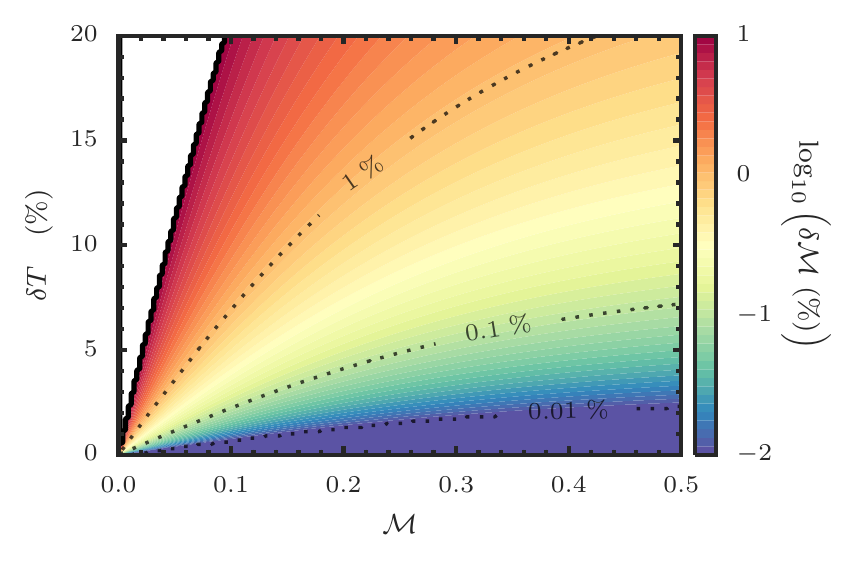}
\caption{Impact of a temperature dispersion on the accuracy of the measurement of \Mach{}. The colouring shows how well an input \Mach{} value can be recovered from a line profile that is the summation of lines at differing temperatures described by $\delta T$.}
\label{fig:temperaturebias}
\end{figure}
}

\newcommand{\FigurePrecision}{
\begin{figure*}
\includegraphics{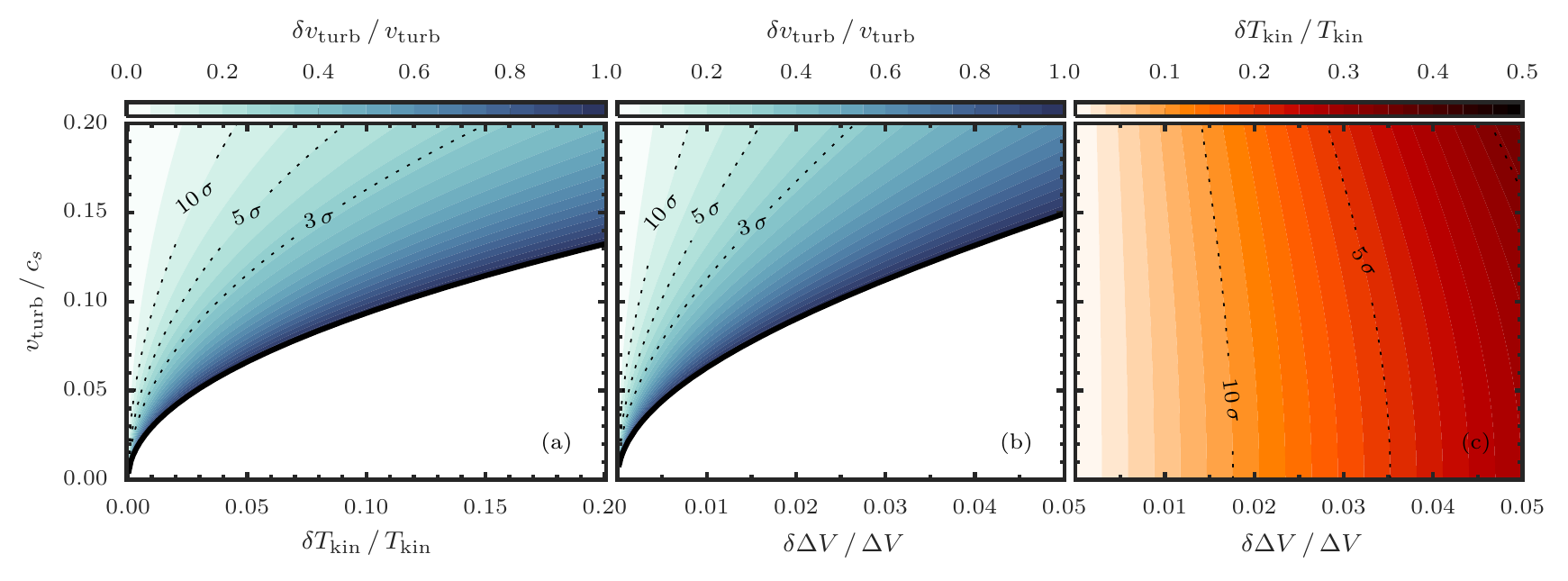}
\caption{Limitations in determining \vturb{} and \Tkin{} from the two direct methods. Panels (a) and (b) show the
relative error in \vturb{} for a given (\vturb{} $/\, c_s$) as a function of the temperature (a) and linewidth (b) corresponding to the direct method with a known excitation temperature and the co-spatial method respectively. Panel (c) shows the relative error in \Tkin{} from the co-spatial method as a function of relative error in linewidth. The left panel does not take into account errors in the linewidth. The dashed contour lines show 10, 5 and 3~$\sigma$ limits.}
\label{fig:limits}
\end{figure*}
}

\newcommand{\FigureTemperatureComparison}{
\begin{figure}
\includegraphics{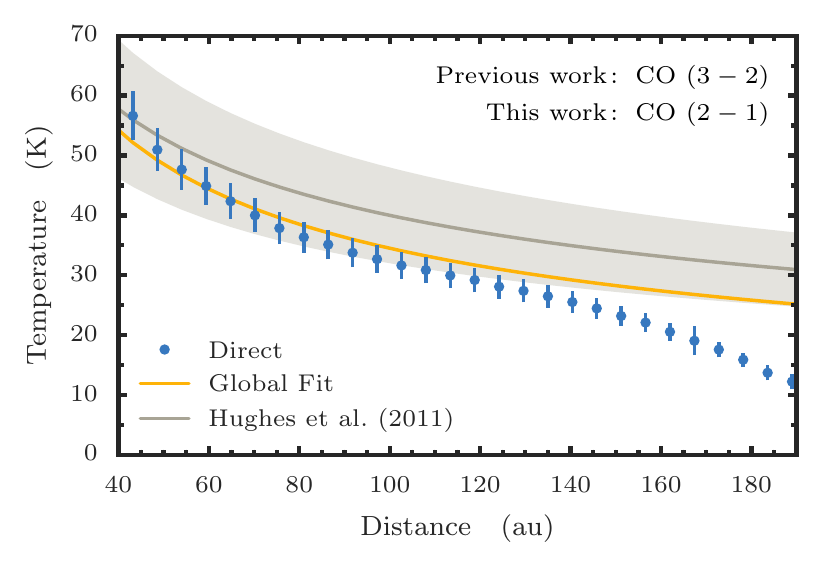}
\caption{Comparing the temperature structure of our fits and that of \citet{Hughes_ea_2011} shown in gray. \citet{Hughes_ea_2011} observed CO (3-2) with the SMA with a flux calibration of $\sim$~20\%. Work from this paper uses CO (2-1) with ALMA which has a flux calibration of $\approx$~7\%.}
\label{fig:temperature_comparisons}
\end{figure}
}

\newcommand{\FigureSpectra}{
\begin{figure*}
\centering
\includegraphics{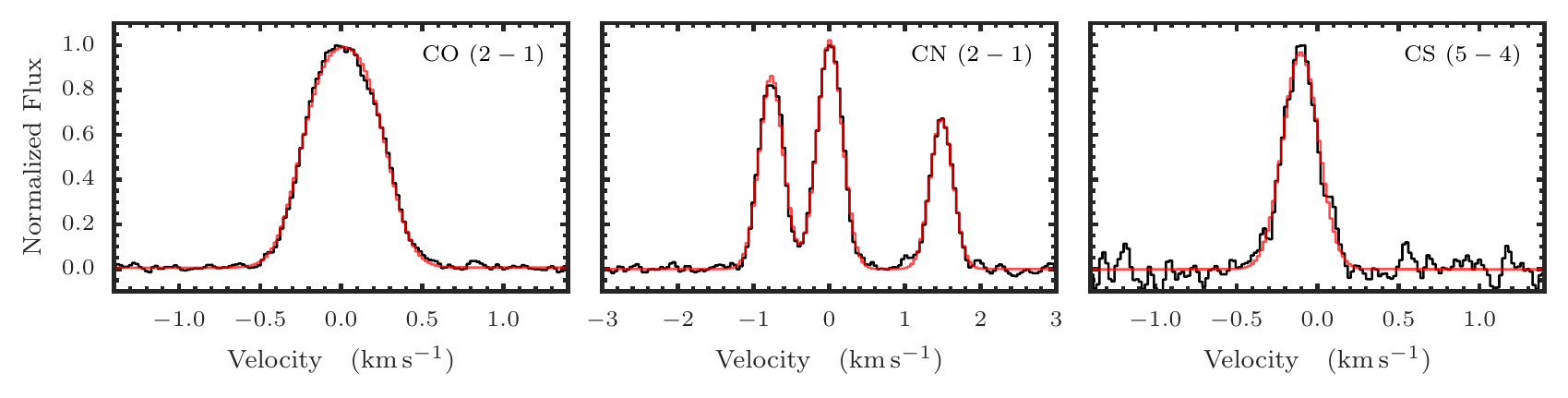}
\caption{Example spectra of CO (left), CN (centre) and CS (right). All spectra are from a pixel $1\arcsec$ from the centre. This example only shows three hyperfine components for CN. The red lines show example fits to the data, CO being an optically broadened Gaussian, CN the ensemble of Gaussians, one for each hyperfine component, and CS a pure, optically thin Gaussian. Note that the spectra are Nyquist sampled in velocity.}
\label{fig:example_spectra}
\end{figure*}
}

\newcommand{\FigureMaps}{
\begin{figure*}
\centering
\includegraphics{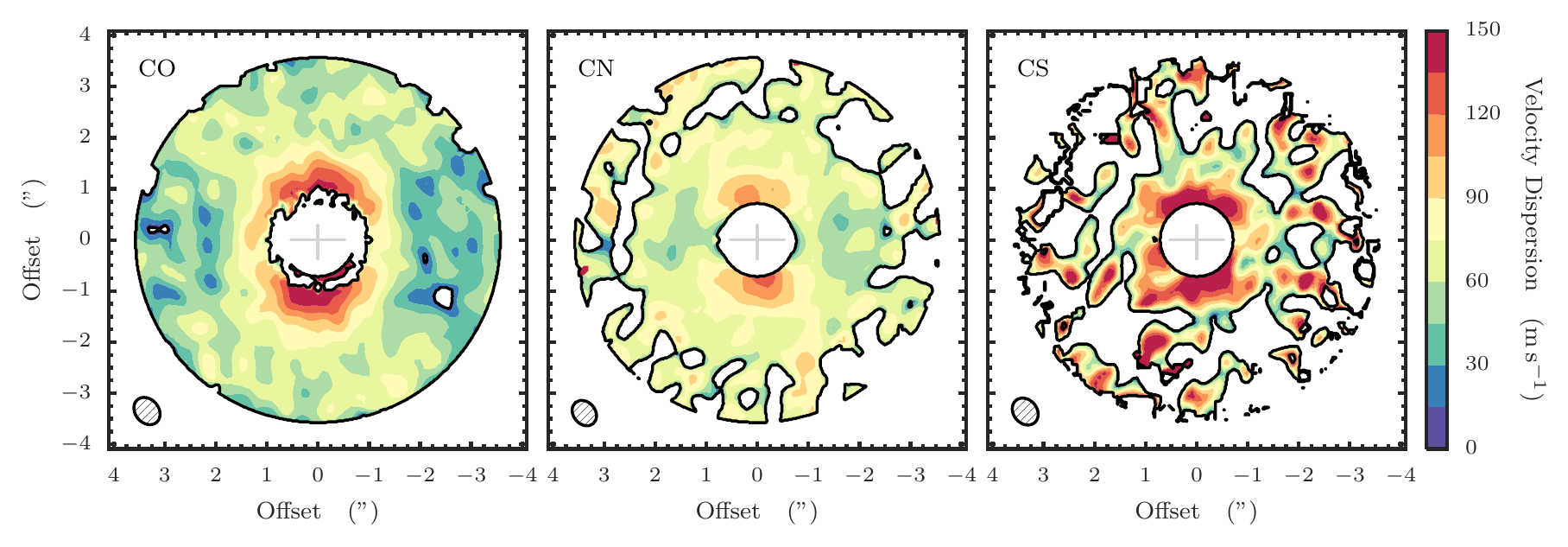}
\caption{2D distribution of \vturb{} for all three lines: CO (left), CN (centre) and the combination of CN and CS assuming co-spatiality, therefore sharing the same \Tex{} and \vturb{} values (right). The values are masked outside 180~au and within 40~au. The beam size is shown in the bottom left corner for each line and the major and minor axes denoted by the central cross, aligned with the x- and y-axes respectively. At the distance of TW~Hya, $1\arcsec \approx 54$~au. The azimuthal asymmetry see in the inner disk is an artefact of a purely radial subtraction of the beam-smearing component discussed in Section~\ref{sec:beamsmear}.}
\label{fig:turbulence_2D}
\end{figure*}
}

\newcommand{\FigureLinewidths}{
\begin{figure*}
\includegraphics{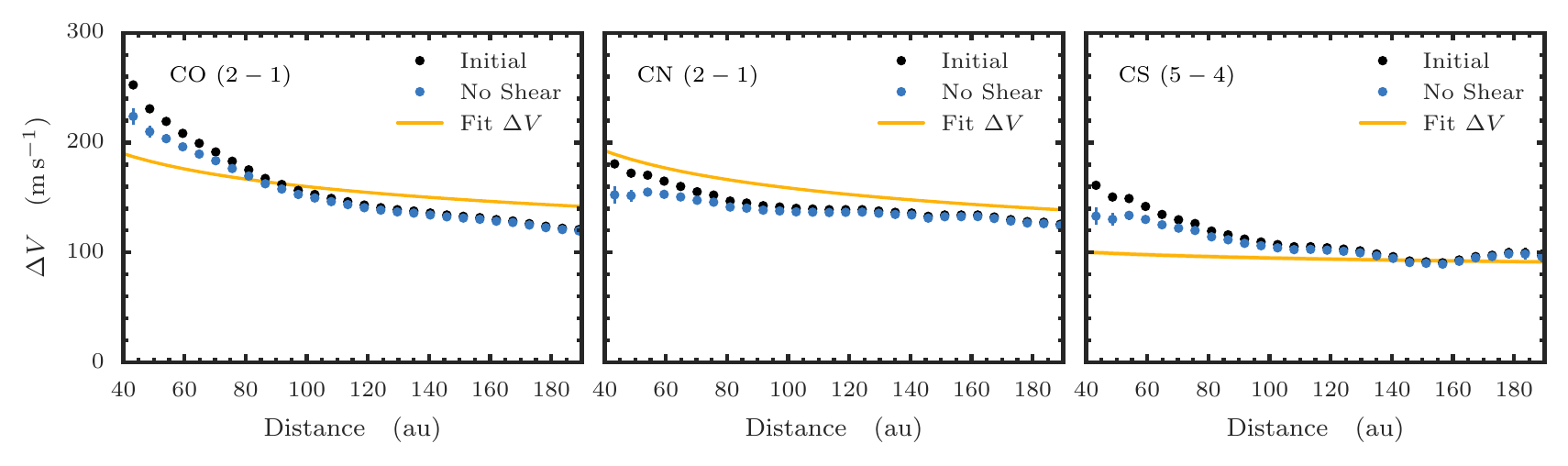}
\caption{Measured (black) and corrected for the Keplerian shear component (blue) linewidths for CO (left), CN (centre) and CS (right). All three lines are subject to the same Keplerian shear component. The uncertainties include an uncertainty on the inclination of the disk of $\pm 2\degr$ as described in Eqn.~\ref{eq:v_kep}. The solid yellow lines show the linewidths from the global fit.}
\label{fig:linewidths}
\end{figure*}
}

\newcommand{\FigureScaledLinewidths}{
\begin{figure}
\includegraphics{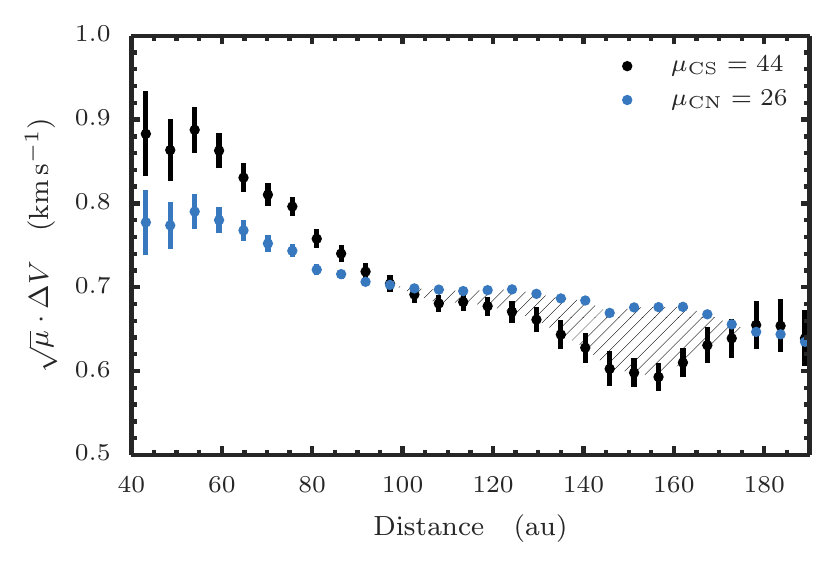}
\caption{Comparing the $\sqrt{\mu}$ scaled linewidths for CN (blue) and CS (black). The region where the scaled linewidth of CS drops below the values for CN, $100 \lesssim r \lesssim 170$~au, shown by the dashes, is where the co-spatial assumption can not be true. Errorbars show $1\sigma$ uncertainties on the mean.}
\label{fig:scaledlinewidths}
\end{figure}
}

\newcommand{\FigureKineticTemps}{
\begin{figure*}
\includegraphics{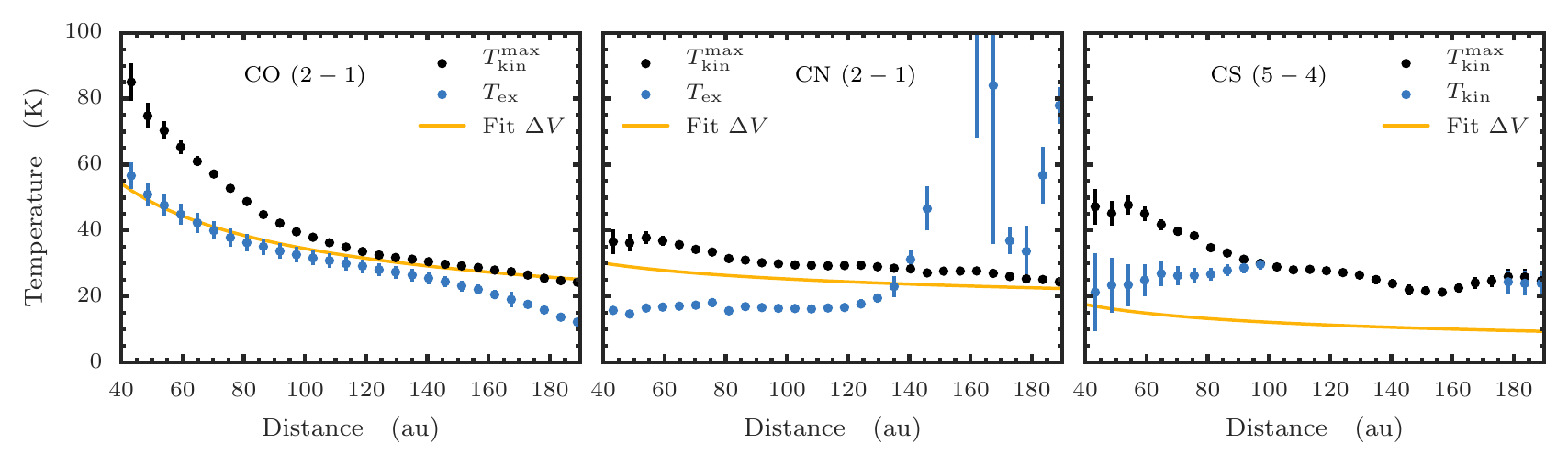}
\caption{Radial profile of the derived $T$ values (in blue) used for calculating the thermal broadening component of the linewidth for CO (left), CN (centre) and CS and CN assuming co-spatiality (right). Note for CO and CN this is \Tex{} while for CS this is \Tkin{}. The black line shows the upper limit \Tkin{} which would fully account for the total linewidth in the absence of turbulent broadening. Outside 140~au the derived \Tkin{} exceeds $T_{\rm kin}^{\rm max}$ for CN and is thus is not considered in further analysis. The black dots in the rightmost panel come from the CS linewidths. Errorbars show $1\sigma$ uncertainties on the mean.} \label{fig:kinetic_temps}
\end{figure*}
}

\newcommand{\FigureDirectTurbulence}{
\begin{figure*}
\includegraphics{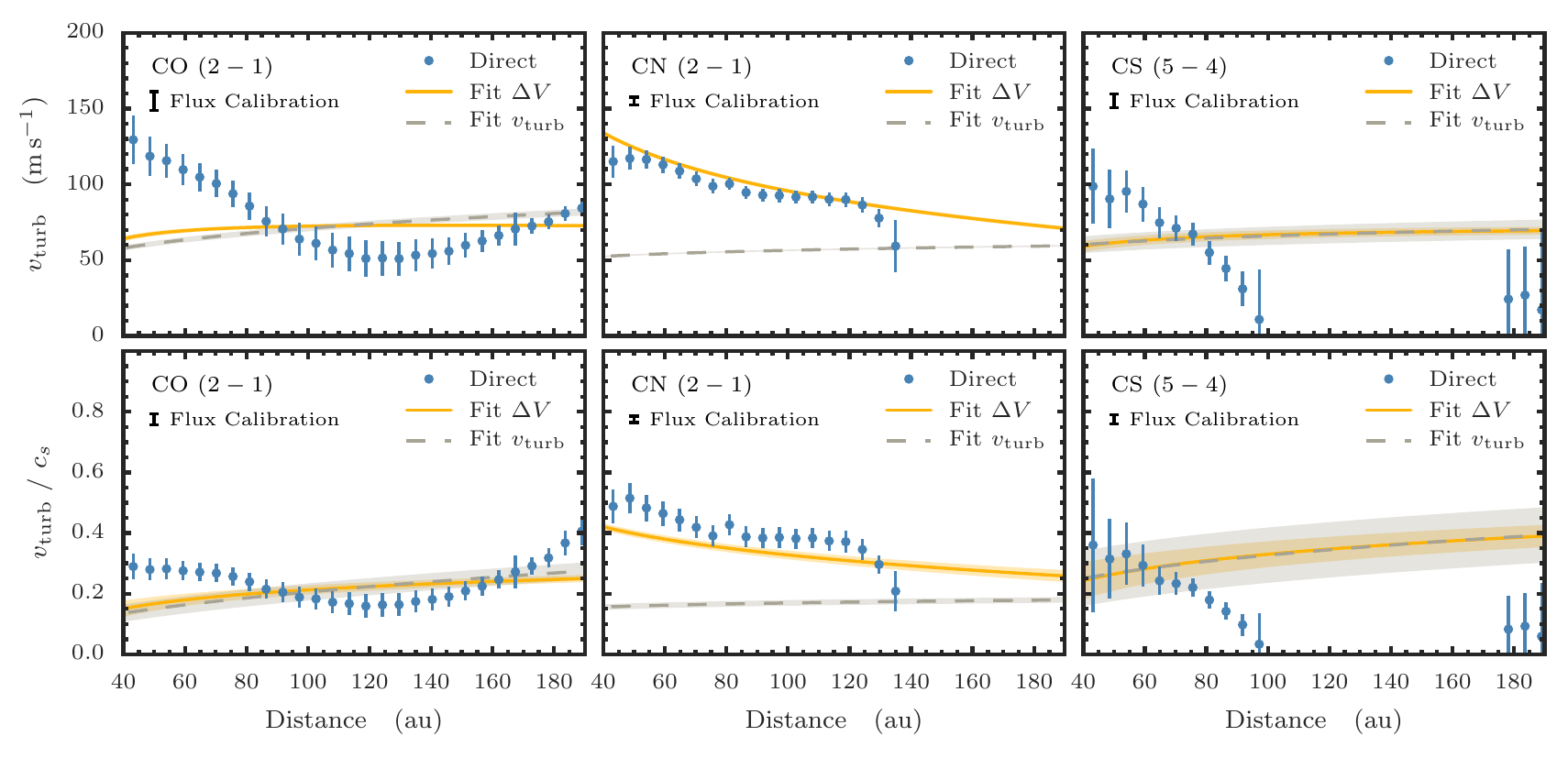}
\caption{Radial profiles of the turbulent width in \vel{}, top row, and as a function of local sound speed, bottom row. The blue dots show the results of the direct method, where the CO and CN lines were assumed to be fully thermalised, and CS to be co-spatial with CN in order to derive a \Tkin{} value. Yellow solid lines show the results from the global fit where the total linewidth was fit for, while dashed gray dashed lines show the global fit where \vturb{} was fit for individually. The 1~$\sigma$ uncertainties are shown as bars on the direct method and shaded regions on the lines. A representative error associated with the flux calibration of 7~\% at 80~au is shown in the top left of all panels.}
\label{fig:direct_turbulent}
\end{figure*}
}

\newcommand{\DiskFitTable}{
\begin{table*}[ht]
\centering
\footnotesize
\doublespacing
\begin{tabular}{ccccccccc} \hline\hline
Line   & $V_{100} \sin(i)$ & $e_v$               & $\Delta V$      & $e_{\Delta V}$    & $T_{100}$        & $e_T$               & $v_{\rm turb}$ & $e_{v_{\rm turb}}$ \\ 
         & (m\,s$^{-1}$)   & (-)                 & (m\,s$^{-1}$)   & (-)               & (K)              & (-)                 & (m\,s$^{-1}$)  & (-)                \\ \hline
\multicolumn{9}{c}{\sc Fitting for a turbulent linewidth component} \\ \hline      
CO J=2-1 & $262.7 \pm 0.2$ & $0.530 \pm 0.001$   & -               & -                 & 35.4 $\pm$ 0.2   & 0.464 $\pm$ 0.001   & 71 $\pm$ 2     & $-0.22 \pm 0.01$   \\
CN N=2-1 & $258.9 \pm 0.6$ & $0.564 \pm 0.002$   & -               & -                 & 33.0 $\pm$ 0.2   & 0.02 $\pm$ 0.04     & 56.5 $\pm$ 0.5 & $-0.08 \pm 0.02$   \\
CS J=5-4 & $261.0 \pm 0.8$ & $0.53 \pm 0.01$     & -               & -                 & 12.1 $\pm$ 0.2   & 0.38 $\pm$ 0.07     & 66 $\pm$ 6     & $-0.10 \pm 0.03$   \\ \hline
\multicolumn{9}{c}{\sc Fitting for a total linewidth} \\ \hline         
CO J=2-1 & $262.7 \pm 0.2$ & $0.535 \pm 0.001$   & $160.0 \pm 0.5$ & $0.187 \pm 0.001$ & 35.51 $\pm$ 0.09 & 0.492 $\pm$ 0.001   & -              & -                  \\
CN N=2-1 & $252.9 \pm 0.4$ & $0.532 \pm 0.007$   & $158.8 \pm 0.8$ & $0.210 \pm 0.003$ & 25.3 $\pm$ 0.2   & 0.19 $\pm$ 0.04     & -              & -                  \\
CS J=5-4 & $261.0 \pm 0.8$ & $0.53 \pm 0.02$     & $95 \pm 2$      & $0.06 \pm 0.01$   & 12.16 $\pm$ 0.08 & 0.40 $\pm$ 0.07     & -              & -                  \\ \hline
\end{tabular}
\singlespacing
\caption{Results of the parametric model fitting. $V_{100}\sin(i)$ is the projected rotation velocity, $\Delta V$ is the total linewidth, $T_{100}$ is the excitation temperature, and \vturb{} the turbulent velocity dispersion, all at 100~au and each with their corresponding exponent. The parameters not fit for were calculated using Eqn.~\ref{eq:linewidth}. Note that for $M_{\star} = 0.69~M_{\sun}$, the measured $V_{100}\sin(i)$ indicates $i=5.96\pm0.03$. \label{tab:disk_params}}
\end{table*}
}

\newcommand{\CNVelocitiesTable}{
\begin{table*}[ht]
\centering
\onehalfspacing
\caption{New Frequencies for CN N=2-1 transitions}
\begin{tabular}{cccc}
\hline \hline 
Old Frequency & New Frequency & Offset & \multirow{2}{*}{Transition} \\
(MHz) & (MHz) & (MHz) & \\ \hline
 226659.5584 & 226659.564 & +0.008 & CN N=2-1 J=3/2-1/2, F=5/2-3/2 \\ 
 226663.6928 & 226663.694 & +0.001 & CN N=2-1 J=3/2-1/2, F=1/2-1/2 \\ 
 226679.3114 & 226679.331 & +0.020 & CN N=2-1 J=3/2-1/2, F=3/2-1/2 \\ 
 226874.1908 & 226874.191 & 0.000  & CN N=2-1 J=5/2-3/2, F=5/2-3/2 \\
 226874.7813 & 226874.781 & [0]    & CN N=2-1 J=5/2-3/2, F=7/2-5/2 \\ 
 226875.8960 & 226875.896 & 0.000  & CN N=2-1 J=5/2-3/2, F=3/2-1/2 \\
 226887.4202 & 226887.403 & -0.017 & CN N=2-1 J=5/2-3/2, F=3/2-3/2 \\ 
 226892.1280 & 226892.128 & 0.000  & CN N=2-1 J=5/2-3/2, F=5/2-5/2 \\
 226905.3574 & 226905.353 & -0.004 & CN N=2-1 J=5/2-3/2, F=3/2-5/2 \\  \hline
\end{tabular}
\singlespacing
\tablefoot{CN N=2-1 line frequencies were measured in laboratory by \citet{Skatrud_ea_1983}; Values in Col.1 are the fitted values from the CDMS Database \citep{CDMS_2001}. Col.2 indicates the values we derived from our spectra.}
\label{tab:cn21}
\end{table*}
}

\begin{document}

   \title{Measuring Turbulence in TW~Hya  with ALMA:\\Methods and Limitations}
   \author{R. Teague
          \inst{1}
          \and
          S. Guilloteau
          \inst{2,3}
          \and
          D. Semenov
          \inst{1}
          \and
          Th. Henning
          \inst{1}
          \and
          A. Dutrey
          \inst{2,3}
          \and
          V. Pi\'etu
          \inst{4}
          \and
          T. Birnstiel
          \inst{1}
          \and\\
          E. Chapillon
          \inst{2,3,4}
          \and
          D. Hollenbach
          \inst{5}
          \and
          U. Gorti
          \inst{5,6}
          }

   \institute{Max-Planck-Institut f\"{u}r Astronomie, K\"{o}nigstuhl 17, 69117 Heidelberg, Germany\\
              \email{teague@mpia.de}
              \and
              Univ. Bordeaux, LAB, UMR 5804, 33270 Floirac, France
              \and
              CNRS, LAB, UMR 5804, 33270 Floirac, France
              \and
              IRAM, 300 rue de la Piscine, Domaine Universitaire, F-38406 Saint Martin d'H\'{e}res, France
              \and
              SETI Institute, 189 Bernardo Avenue, Mountain View, CA 94043, USA
              \and
              NASA Ames Research Center, Moffett Field, CA, USA}

   \date{Accepted 30/05/16}

\abstract
{}
{To obtain a spatially resolved measurement of velocity dispersions in the disk of TW~Hya.}
{We obtain high spatial and spectral resolution images of the CO J=2-1, CN N=2-1 and CS J=5-4 emission with ALMA in Cycle~2. The radial distribution of the turbulent broadening is derived with three approaches: two `direct' and one modelling. The first requires a single transition and derives \Tex{} directly from the line profile, yielding a \vturb{}. The second assumes two different molecules are co-spatial thus their relative linewidths allow for a calculation of \Tkin{} and \vturb{}. Finally we fit a parametric disk model where physical properties of the disk are described by power laws, to compare our `direct' methods with previous values.}
{The two direct methods were limited to the outer $r > 40$~au disk due to beam smear. The direct method found \vturb{} ranging from $\approx$~\vel{130} at 40~au, dropping to $\approx$~\vel{50} in the outer disk, qualitatively recovered with the parametric model fitting. This corresponds to roughly $0.2 - 0.4~c_s$. CN was found to exhibit strong non-LTE effects outside $r \approx 140$~au, so \vturb{} was limited to within this radius. The assumption that CN and CS are co-spatial is consistent with observed linewidths only within $r \lesssim 100$~au, within which \vturb{} was found to drop from \vel{100} ($\approx~0.4~c_s$) to nothing at 100~au. The parametric model yielded a near constant \vel{50} for CS ($0.2 - 0.4~c_s$). We demonstrate that absolute flux calibration is and will be the limiting factor in all studies of turbulence using a single molecule.}
{The magnitude of the dispersion is comparable with or below that predicted by the magneto-rotational instability theory. A more precise comparison would require to reach an absolute calibration precision of order 3\%, or to find a suitable combination of light and heavy molecules which are co-located in the disk.}
   \maketitle

\section{Introduction}
\label{sec:introduction}

Turbulent motions underpin the entire evolution of a protoplanetary disk. Foremost, turbulence determines the bulk gas viscosity and hence regulates the angular momentum transport and accretion in disks \citep{Shakura_ea_1973,Pringle81}. Secondly, turbulence is a key factor for dust evolution and transport in disks \citep{Testi_ea_2014,Henning_Meeus_2011}. However, until recently, observational constraints on the level of disk turbulence were extremely challenging to obtain and hence scarce. With the advent of the Atacama Large Millimeter / submillimeter Array (ALMA), we have access for the first time to the high sensitivity, spectral resolution and  angular resolution observations which are needed to directly measure turbulent velocities in disks.

\FigureSpectra

Accurate determination of the turbulent velocity dispersion from line broadening requires a good understanding of the other components which contribute to the line width, namely bulk motions of the gas, thermal broadening and, in the case of a highly optically thick line, broadening due to the line opacity. All previous measurements of \vturb{} have revolved around the fitting of a parametric model in order to extract a disk-averaged turbulent broadening value. The derived values ranged from very low values of $\la 10-100$~m\,s$^{-1}$ ($\la 0.02 - 0.2$~$c_s$) derived for the TW~Hya and HD~163296 disks, to higher velocities of $\la 100 - 200$~m\,s$^{-1}$ ($\la 0.3-0.5$~$c_s$) for the disks of DM~Tau, MWC~480 and LkCa~15 \citep{Dartois_ea_2003,Pietu_ea_2007,Hughes_ea_2011,Rosenfeld_ea_2012,Flaherty_ea_2015}. With the exception of TW~Hya and HD~163296 \citep{Hughes_ea_2011,Flaherty_ea_2015}, the spectral resolution of the data used to determine these values, of the order $\sim$~\vel{200}, is too coarse to resolve the small expected contribution from turbulent broadening, although \citet{Guilloteau_ea_2012} did correct for this effect when using CS to measure turbulence in DM Tau.

High-quality ALMA Cycle~2 observations of TW~Hya allow us for {\em the first time} to get a direct measure of the line widths and thus spatially resolved turbulent velocity structure. With a near face-on inclination of only $i \approx 7\degr$ \citep{Qi_ea_2004} and as the nearest protoplanetary disk at $d \approx 54$~pc, TW~Hya provides the best opportunity to directly detect turbulent broadening as the impact of Keplerian shear for such face-on disks is minimized compared to more inclined systems.

We present here the first direct measurements of \vturb{} in a protoplanetary disk using the line emission of CO, CN and CS. In Sect.~\ref{sec:observations} we describe our ALMA observations and the data reduction. Section~\ref{sec:vturb} describes the methods we used to extract \vturb{}: two direct methods relying on a measure of the line widths and a more commonly used fit of a parametric model. Discussion in Sect.~\ref{sec:discussion} follows.


\section{Observations}
\label{sec:observations}

The observations were performed using ALMA on May 13, 2015 under excellent weather conditions (Cycle 2, 2013.1.00387.S). The receivers were tuned to cover CO J=(2-1), CS J=(5-4) and all strong hyperfine components of CN N=(2-1) simultaneously. The correlator was configured to deliver very high spectral resolution, with a channel spacing of 15~kHz (and an effective velocity resolution of \vel{40}) for the CO J=(2-1) and CS J=(5-4) lines, and 30~kHz (\vel{80}) for the CN N=(2-1) transition.

Data was calibrated using the standard ALMA calibration script in the \texttt{CASA} software package\footnote{\url{http://casa.nrao.edu/}}. The calibrated data were regridded in velocity to the LSR frame, and exported through UVFITS format to the \texttt{GILDAS}\footnote{\url{http://www.iram.fr/IRAMFR/GILDAS}} package for imaging and data analysis. Self-calibration was performed on the continuum data, and the phase solution applied to all spectral line data. With robust weighting, the $uv$ coverage, which have baselines between 21 and 550~m, provided by the $\sim 34$ antennas yields a beamsize of $0.50\arcsec \times 0.42\arcsec$ at position angle of $80\degr$.The absolute flux calibration was referred to Ganymede. The derived flux for our amplitude and phase calibrator, J1037-2934, was 0.72 Jy at 228 GHz at the time of the observations, with a spectral index $\alpha = -0.54$, while the ALMA flux archive indicated a flux of $0.72 \pm 0.05$ Jy between April 14$^{\rm th}$ and April 25$^{\rm th}$. We hence estimate that the calibration uncertainty is about 7\%.

After deconvolution and primary beam correction, the data cubes were imported into \texttt{CLASS} for further analysis, in particular line profile fits including the hyperfine structure for CN lines. For the azimuthal average, each spectrum was shifted in velocity from its local projected Keplerian velocity before averaging. We used for this the best fit Keplerian model assuming a stellar mass of $0.69~M_{\sun}$ and $i = 7\degr$, see Section.~\ref{sec:method3}. 

All three emission lines show azimuthal symmetry within the noise justifying our choice to azimuthally average the data. CO emission looks identical to previous studies (for example \citet{Qi_ea_2013}), while integrated intensity plots for CN, including all hyperfine components, and CS are shown in Appendix~\ref{sec:appobservations}. Sample spectra illustrating the very high signal to noise obtained in CO and CN, and the noisier CS data, are given in Fig.~\ref{fig:example_spectra} and a gallery of azimuthally averaged spectra at different radial locations can be found in Appendix~\ref{sec:appobservations}. Finally, examples of the full compliment of CN hyperfine components are found in Fig.~\ref{fig:aaallspectra}.

\section{Disentangling Turbulent Velocity Dispersions}
\label{sec:vturb}

\FigureLinewidths

Turbulent motions within a gas manifest themselves as a velocity dispersion along the line of sight, broadening the width of the emission (or absorption) line. This broadening term acts in tandem with thermal broadening, a contribution typically an order of magnitude larger than the turbulent width. Additionally, the Keplerian shear across the beam will broaden the observed emission lines. This effect is the most dominant in the inner disk and for highly inclined disks, making TW~Hya an ideal source as this effect is minimized.

In the following section we discuss three methods to extract \vturb{}, the turbulent velocity dispersion: two `direct' and one `parametric' approach, and apply each to TW~Hya.

\subsection{Line width measurements}
\label{sec:linewidthmeasurements}

Physical parameters were extracted from the line profiles at each pixel in the image and for an azimuthal average. CO is highly optically thick and displays a saturated core meaning the line profile deviates strongly from an optically thin Gaussian (see left panel of Fig.~\ref{fig:example_spectra}). By fitting a line profile of the form,

\begin{equation}
I_v = \Big(J_{\nu}(T_{\rm ex}) - J_{\nu}(T_{\rm bg}) \Big) \cdot \left( 1 -  \exp \left[ -\tau \exp \left\{ -\frac{(v - v_0)^2}{\Delta V^2}\right\} \right] \right),
\end{equation}

\noindent \new{where $T_{\rm bg} = 2.75$~K,} we are able to obtain the line full-width at half maximum, FWHM, line center $v_0$ and, if the line is sufficiently optically thick, \Tex{} and $\tau$ (otherwise only the product is constrained).

Under the assumption that all hyperfine components arise from the same region in the disk and that the main component is optically thick, the relative intensities of the CN hyperfine components yield an optical depth and \Tex{}. Using the `hyperfine' mode in \texttt{CLASS} the hyperfine components were simultaneously fit with Gaussian profiles. It was found that the recommended spacing of hyperfine components were systematically biased across the disk, suggesting that the recommended offset values were incorrect. Fitting for the relative positions of each component allowed for a better determination of their spacing to $\approx$~\vel{1}. The adopted frequencies are given in Table~\ref{tab:cn21}.

Finally, the CS emission was well fit by an optically thin Gaussian, from which the linewidth and line center were able to be extracted accurately. However with only a single transition the degeneracy between \Tex{} and $\tau$ could not be broken so we remain ignorant on the local temperature.

The linewidths are sufficiently well sampled with spectral resolutions of \vel{40} for CO and CN and \vel{37} for CS such that sampling effects are negligible for our data. Assuming square channels and Gaussian line profiles, we estimate that the bias on the measured $\Delta V$ would be $\approx 2~\%$ for CO and CN and $\approx 3.5~\%$ for CS. Figure~\ref{fig:resolution} shows the impact of the resolution on the determination of $\Delta V$. These biases have been included in the following analysis.

\subsection{Keplerian shear correction}
\label{sec:beamsmear}

In the following `direct' methods we only consider the disk outside 40~au. Within this radius the spectra start to strongly deviate from the assumed Gaussian (in opacity) line profiles, because parts of the disk rotating in opposite directions are smeared in the beam. Given the flux calibration, there is an intrinsic 7\% uncertainty on the peak values of the spectra, thus the \Tex{} values derived for CO and CN have uncertainties of at least 7\%. The impact of this is discussed in Sec.~\ref{sec:discussion}.

To estimate the impact of the artificial broadening due to the beam smear, the physical model of TW~Hya from \citet{Gorti_ea_2011} was used. The model was run through the \texttt{LIME} radiative transfer code \citep{Brinch_ea_2010} for a range of inclinations, $i = \{0\degr,\,5\degr\,,6\degr\,,7\degr\,,8\degr\,,9\degr\}$, assuming no turbulent broadening. We note that the projected velocity is a product of both stellar mass and inclination. Thus, by varying only the inclination we are able to consider uncertainties in both quantities\footnote{The relative error $\delta i / i \approx 0.29$ considered is equivalent to assuming $\delta M_{\star} / M_{\star} = 0.58$. Alternatively, this could be considered as $M_{\star} = 0.6 \pm 0.15~M_{\sun}$ and $i = 7 \pm 1.9~\degr$, well representative of TW~Hya.}.

Following \cite{Rosenfeld_ea_2013}, we account for the height above the midplane in the calculation of the velocity field. Both CO J=(2-1) and C$^{18}$O (2-1) lines were modelled, allowing us to sample both an optically thick and thin case. Using \texttt{CASA}, the model observations were converted to synthetic observations with the same array configuration as the true observations. Differences in the resulting line width at each pixel between an inclined disk and a face on disk were attributed to Keplerian broadening.

At our linear resolution ($\sim 25$ au), the radial distribution of differences in linewidths was well fit by a power law outside of 40~au,

\begin{equation}
\Delta V_{\rm Kep} = \Big(2.6 \pm 0.5\Big) \times \left(\frac{r}{100} \right)^{\,-3.2 \pm 0.1} \,\, {\rm m\,s^{-1}},
\label{eq:v_kep}
\end{equation}

\noindent with $r$ the radial distance in au. Quoted uncertainties are $1\sigma$ and are dominated by an uncertainty in inclination of $\pm 2\degr$. The differences between the $^{12}$CO and C$^{18}$O cases were smaller than these quoted uncertainties.

This component was subtracted from all linewidths prior to further analysis. Figure~\ref{fig:linewidths} shows the measured linewidths (black lines), and the linewidths after the correction for Keplerian shear (blue lines).

\FigureKineticTemps

\subsection{Single Molecule Approach}
\label{sec:method1}

After correcting for the Keplerian shear we assume the linewidth is only a combination of thermal and turbulent broadening. Hence the remaining linewidth can be described as,

\begin{equation}
\Delta V = \sqrt{v_{\rm turb}^2 + \frac{2kT_{\rm kin}}{\mu m_{\rm H}}},
\label{eq:linewidth}
\end{equation}

\noindent where $\mu$ is the molecular mass of the tracer molecule, $m_{\rm H}$ the mass of a hydrogen atom, the kinetic temperature of the molecule \Tkin{} and the linewidth $\Delta V = {\rm FWHM} \,/\, \sqrt{4 \ln 2}$.

For both CO and CN, the line profiles provided \Tex{}, so a conversion to \Tkin{} must be made. Guided by the particle densities in the model of \citet{Gorti_ea_2011} at the region of expected emission for CO and CN, $\ga 10^6-10^7$~cm$^{-3}$, we make the assumption that both CO and CN lines are thermalised so that \Tex{} = \Tkin{} $ = T$. The validity of this assumption is discussed in Sec.~\ref{sec:discussion}. Derived \Tkin{} values for CO and CN are shown by the blue lines in the left two panels of Fig.~\ref{fig:kinetic_temps}. The black lines show $T^\mathrm{max}_\mathrm{kin}$, the maximum kinetic temperature in the absence of any turbulence:

\begin{equation}
 T^\mathrm{max}_\mathrm{kin} = \frac{\mu m_{\rm H}}{2 k} \,\Big(\Delta V\Big)^2.
\end{equation}

\noindent In essence, the residual between these two lines must be accounted for either by turbulent broadening, or sub-thermal excitation, i.e. \Tkin{} $> T$.

Outside of $r \sim 140$~au, CN shows signs of non-LTE effects as the derived \Tex{} is considerably higher than $T_{\rm kin}^{\rm max}$, indicating the presence of weak pumping of the line (see Fig.~\ref{fig:kinetic_temps}). These `supra-thermal' regions are neglected in the remainder of the analysis. The (small) impact of unresolved turbulence and or temperature gradients in the finite beamsize will be discussed Section~\ref{sec:discussion}.

With a known \Tkin{} a simple subtraction of the thermal broadening component leaves \vturb{}. The left two columns of Fig.~\ref{fig:direct_turbulent} show the derived \vturb{} in units of \vel{} in the top panel and as a function of local soundspeed $c_s$ in the bottom panel for CO and CN respectively. Fig.~\ref{fig:turbulence_2D} shows the spatial distribution of \vturb{} (we neglected here the primary beam correction, which only reaches 7\% at the map edge). For the case of CS, the line is essentially optically thin, and we cannot derive an excitation temperature.

\subsection{Co-Spatial Approach}
\label{sec:method2}

Instead of relying on the temperature derived from a single molecule, we can take advantage of molecules with different molecular weights to separate the thermal and turbulent broadening, assuming the lines from these molecules emit from the same location in the disk. Under this assumption the total linewidths would be tracing the same \vturb{} and \Tkin{}. Solving Equation~\ref{eq:linewidth} simultaneously for two molecules, $A$ and $B$ with respective molecular masses,  $\mu_{\rm A}$ and $\mu_{\rm B}$ where $\mu_{\rm A} < \mu_{\rm B}$, and total linewidths, $\Delta V_{\rm A}$ and $\Delta V_{\rm B}$, we find,

\begin{align}
T_{\rm kin} &= \frac{m_{\rm H}}{2k} \frac{\mu_{\rm A} \, \mu_{\rm B}}{\mu_{\rm B} - \mu_{\rm A}} \, \Big( \Delta V_{\rm A}^2 - \Delta V_{\rm B}^2 \Big), \label{eq:simtkin}\\
v_{\rm turb} &= \sqrt{\frac{\mu_{\rm B} \Delta V_{\rm B}^2 - \mu_{\rm A} \Delta V_{\rm A}^2}{\mu_{\rm B} - \mu_{\rm A}}} \label{eq:simvturb}.
\end{align}

\noindent This method does not make any assumption about the excitation temperature of the observed transitions, but relies only on the measured linewidths and the co-spatiality of the emitting regions.

Among the observed molecules, CO may only trace a narrow layer because of its high optical depth. However, one would expect the optically thin CN and CS to trace a larger vertical region. Both CN and CS would freeze-out at a similar temperature so the bottom of their respective molecular layers would be relatively coincident, thus potentially trace the same region in the disk. Hence we choose to apply this method to the two lines of CN and CS.

The right most panel of Fig.~\ref{fig:kinetic_temps} shows the \Tkin{} (blue line) derived from CN and CS, in comparison to $T^\mathrm{max}_\mathrm{kin}$, the maximum \Tkin{} derived from the CS linewidth (black). Radial profiles of \vturb{} derived from CN and CS are shown in the right column of Fig.~\ref{fig:direct_turbulent}, in \vel{} (top) and as a function of $c_s$ (bottom).

Gaps in \Tkin{} and \vturb{} correspond to where the $\mu$-scaled linewidth of CS is less than the $\mu$-scaled linewidth of CN (see Fig.~\ref{fig:scaledlinewidths}). In this situation there is no solution to Eqs.~\ref{eq:simtkin} and \ref{eq:simvturb}, thus the assumption of CN and CS being cospatial fails.

\subsection{Parametric Model Fitting}
\label{sec:method3}

\DiskFitTable

The above direct methods require a proper correction of the Keplerian shear, which scales as $\sqrt{M_*} \sin(i)$. For edge-on disks, or when the angular resolution is insufficient to remove the Keplerian shear, our direct technique is unapplicable, and the only available method is to use a parametric model assuming \Tkin{} and the total local linewidth $\Delta V$. A parametric model fit can recover $\Delta V$ with high accuracy independently of the absolute (flux) calibration error. However, the fraction of this width which is due to turbulence depends on the absolute calibration since the thermal line width scales as the square root of the kinetic temperature. 

In the following we give a brief description of the parametric model but refer the reader to \citet{Dartois_ea_2003} and \citet{Pietu_ea_2007} for a thorough model description and fitting methodology. The model assumes a disk physical structure which is described by an ensemble of power laws:

\begin{equation}
A_{\rm r} = A_{100} \times \left( \frac{r}{100}\right)^{-e_{\rm A}},
\end{equation}

\noindent for some physical parameter $A$ and cylindrical distance $r$ in au. A positive $e_a$ means the $A$ parameter \emph{decrease} with radius. The molecule densities follow a Gaussian distribution in $z$, whose scale height $H$ is used as a free parameter (this is equivalent to a uniform abundance in a vertically isothermal disk). This method allows to correct to first order the geometric effects in the projected rotation velocities due to disk thickness. CO was found to sample a much higher layer (larger $H$) than CN or CS which yielded similar values. With this method we fit two models, firstly one used previously in the literature where \vturb{} is described as a radial power-law, and secondly where we fit for the total linewidth, $\Delta V$, then calculate the value of \vturb{} from Equation.~\ref{eq:linewidth}. Note that by fitting for $\Delta V$ we result in a non-power-law description of \vturb{}.

An inclination, position angle and systemic velocity were found that were comparable to literature values: $i \approx 6\degr$, ${\rm PA} \approx 240\degr$ and $V_{\rm LSR} \approx 2.82$~km\,s$^{-1}$. Physical parameters relevant to \vturb{} are found in Table~\ref{tab:disk_params} along with their formal errors. All three molecules yielded a steeper dependance of $e_v$ than a Keplerian profile with $e_v \approx 0.53$. Such a change in projected velocity could either be a projection effect, such as a warp in the disk \citep{Roberge_ea_2005,Rosenfeld_ea_2012}, or gas pressure resulting in non-Keplerian rotational velocities for the gas \citep{Rosenfeld_ea_2013}. To account for such an exponent with a warp, $i$ needs to change by $\approx 1\degr$ between 40 and 180~au. Thus, while this non-Keplerian bulk motion was not considered explicitly in the removal of the Keplerian shear, the range of inclinations considered, $7 \pm 2\degr$, sufficiently accounts for such a deviation. Further analysis of this is beyond the scope of this paper.

As with the two direct methods, it was assumed all lines were fully thermalised so that the excitation temperature recovered the full thermal width of the line. A comparison of the total linewidths, temperature profiles and turbulent components are shown in yellow solid lines in Figs.~\ref{fig:linewidths},~\ref{fig:kinetic_temps} and \ref{fig:direct_turbulent} respectively.

\section{Results and Discussion}
\label{sec:discussion}

In the previous section we have described three approaches we have used to measured \vturb{} in TW~Hya. In the following section we compare the methods and discuss their limitations with a view to improving them.

\subsection{Temperature Structure}

Thermal and turbulent broadening are very degenerate and so a precise determination of the temperature structure is pre-requisite to deriving the level of turbulent broadening. Both direct and parametric methods yield comparable temperatures for CO and CN, as shown in Fig.~\ref{fig:kinetic_temps}, however find largely different values for \vturb{}, demonstrating the sensitivity of \vturb{} to the assumed temperature structure.

Excitation temperatures derived from the parametric modelling approach yielded warmer temperatures for CO than CN, in turn warmer than CS with $T_{100}$ = $35.4 \pm 0.2$~K, $25.3 \pm 0.2$~K and $12.2 \pm 0.1$~K respectively when fitting for a total linewidth (see Table~\ref{tab:disk_params}), a trend that was also seen in the direct methods. These values suggest that the emission from each molecule arises from a different height above the midplane in the disk and therefore could be used to trace the vertical structure of \vturb{}.

In the single molecule analyis, either direct or parametric, it was assumed that for both CO and CN, \Tex{} = \Tkin{}, that is, they are both in local thermal equilibrium (LTE). This assumption was guided by the model of \citet{Gorti_ea_2011} which has particle densities of $\ga 10^6-10^7$~cm$^{-3}$ where we believe the molecular emission of CO and CN to arise from. This is sufficient to thermalise the CO line. Given that \Tkin{} $\geq$ \Tex{}, except from the extremely rare case of supra-thermal excitation, the above analysis yielded a lower limit to \Tkin{}, therefore an upper limit to \vturb{}. However, for CN, we have clear evidence for supra-thermal excitation beyond 130 au. A detailed discussion of this issue is beyond the scope of this article. In the future, with multiple transitions it is possible to use the relative intensities of the transitions to guide modeling of the excitation conditions traced by the molecule, thereby yielding a more accurate scaling of \Tex{} to \Tkin{}.

The co-spatial assumption for CN and CS clearly fails in certain regions of the disk where there is no solution to Eqs.~\ref{eq:simtkin} and \ref{eq:simvturb}. Indeed, the temperatures derived from the parametric modeling yield considerably different temperatures for both CN and CS (see Table~\ref{tab:disk_params}), suggesting that this co-spatial assumption fails across the entire disk. Chemical models suggest that CN is present mostly in the photon-dominated layer, higher above the disk plane than CS (although S-bearing molecules are poorly predicted by chemical models, see \cite{Dutrey_ea_2011}). The non-thermalization of the CN N=2-1 line that we observe beyond 130~au also supports the presence of CN relatively high above the disk plane. The accuracy of this assumption can be tested, as well as searching for other co-spatial molecular tracers, with the observation of edge-on disks where the `molecular layers' can be spatially resolved.

Measurements of temperature will be sensitive to temperature gradients along the line of sight, both vertically and radially. Radial gradients will prove more of an issue than vertical as molecular emission will arise predominantly from a relatively thin vertical region, so we expect only a small dispersion vertically in temperature. With the temperature profiles discussed in Section.~\ref{sec:method3}, we estimate that the radial average dispersion across the beam is $\delta T_{\rm beam} \lesssim 5~\%$ outside 40~au for all three lines with a maximum of $\sim 10~\%$ for the very inner regions.

\FigureTemperatureBias

To understand the impact of this on the subsequent derivation of \vturb{}, we consider a two-zone model. We take two regions of differing temperature, but sharing the same turbulent velocity described by a Mach number, $\mathcal{M}_{\rm true} = v_{\rm turb} \, / \sqrt{2} \, c_s$ and the same optical depth. We measure a temperature and linewidth using a Gaussian line profile of the resulting combined line profile and derive a Mach number, $\mathcal{M}_{\rm obs}$. With this method we are able to explore how accurately $\mathcal{M}_{\rm obs}$ can recover $\mathcal{M}_{\rm true}$ with a given temperature dispersion. Figure~\ref{fig:temperaturebias} shows the relative error on \Mach{}, $\delta \mathcal{M}$, as a funtion of $\mathcal{M}_{\rm true}$ and temperature dispersion $\delta T$, assuming the main temperature is 30~K. Taking the temperature dispersions across the beam of 10~\%, we find an uncertainty of $\lesssim 1~\%$ for \Mach{}. This suggests that our determination of \vturb{} is not biased by the expected line of sight gradients in temperature and turbulent width.

\subsection{Turbulent Velocity Dispersions}

With an assumed thermal structure, the turbulent broadening component was considered the residual linewidth not accounted for by thermal broadening or beam smear. Resulting values of \vturb{} are compared in Fig.~\ref{fig:direct_turbulent} and \ref{fig:turbulence_2D}. All three methods yielded values of \vturb{} that ranged from $\sim 50 - 150 \, {\rm m\,s^{-1}}$ corresponding to the range $\sim 0.2 - 0.4~c_s$, however exhibit different radial profiles. The azimuthal structure seen near the centre of the disk in all panels of Fig.~\ref{fig:turbulence_2D} is due to the azimuthal-independent subtraction of beam smearing used in Section~\ref{sec:linewidthmeasurements}.

\FigureDirectTurbulence

\subsubsection{Single Molecule Approach}

CO and CN emission allowed for a `single molecule approach' as described in Section~\ref{sec:method1}. CO yielded values of \vturb{} for $40 \lesssim r \lesssim 190$~au while CN was limited to $40 < r \lesssim 130$~au because of the potential non-LTE effects described in the previous section. Both molecules displayed a decreasing \vturb{} with radius, although CO has a slight increase in the other edges. As a fraction of $c_s$, both molecules ranged within $\sim 0.2 - 0.4~c_s$, however for CO this was found to increase with radius while CN decreased.

\subsubsection{Co-Spatial Approach}

Assuming CN and CS are co-spatial, we find \vturb{} values ranging from \vturb{}~$\leq$~\vel{100} or \vturb{}~$\leq~0.4~c_s$, comparable to the range found for CO and CN individually.

This method, however, is limited by the validity that CN and CS are co-spatial. Indeed, the assumption fails absolutely between $100 \lesssim r \lesssim 180$~au where the linewidth measurements do not allow for a solution of Eqns.~\ref{eq:simtkin} and \ref{eq:simvturb} to be found. This is more clearly seen in Fig.~\ref{fig:scaledlinewidths} which shows the linewidths of CN and CS scaled by $\sqrt{\mu}$ where $\mu = 26$ for CN and $\mu = 44$ for CS. In the region where no solution is found the scaled linewidth for CS is less than that of CN. Despite this limitation, this suggests the utility of another method of determining both \Tkin{} and \vturb{}.

\subsubsection{Parametric Model Fitting}

All previous measurements of \vturb{} have relied on fitting a power-law model of a disk to the observations \citep{Dartois_ea_2003,Pietu_ea_2007,Hughes_ea_2011,Guilloteau_ea_2012,Rosenfeld_ea_2012,Flaherty_ea_2015}, so this allows for a direct comparison to previous results in the literature. In addition, with data with reduced spatial and spectral resolution, the `direct' methods will not be possible so it is important to validate the parametric modelling approach.

We have described two models which were fit to the data with the results shown in Table~\ref{tab:disk_params}. Both include the excitation temperature as a radial power-law, however for one we assume the total linewidth is a power-law, while for the other we assume \vturb{} is a power-law. Accordingly, the parameter not fit for is \emph{not} a power law, but rather derived through Eqn.~\ref{eq:linewidth}. Typically with high spectral and spatial resolution, the data only allows for the second method. A comparison between the models are shown in Fig.~\ref{fig:direct_turbulent} where the yellow solid line shows the case where $\Delta V$, the total linewidth, was assumed a power-law, and the dashed gray lines are where \vturb{} was assumed a power-law. All three molecules display similar ranges of \vturb{}, $\sim 50 - 150 \, {\rm m\,s^{-1}}$  ($\sim 0.1 - 0.4~c_s$) as the direct methods. 

For both CO and CS the two parametric models yield similar results, however the second, where \vturb{} is fit for, has larger uncertainties. Both molecules have a slightly increasing \vturb{} with radius $e_{v_{\rm turb}} \approx -0.22$ and $-0.1$ respectively around \vel{60}. CN, on the other hand, shows a distinct dichotomy between the two due to the different temperature profiles derived for the two methods (see Table~\ref{tab:disk_params}). As mentioned in the previous section, CN displays non-LTE effects which the LTE parametric model may struggle to fit. 

A limiting feature of such parametric model fitting is showcased by the results of CO (left column of Fig.~\ref{fig:direct_turbulent}). If the physical properties of the disk vary from a power-law description, the model will fail to fit this and may be driven to the best `average' description. For example, while the power-law method recovers \vturb{} for CO for $r \gtrsim 100$~au, inside of this radius the two derived \vturb{} vales, one directly and one from model fitting, can deviate by up to a factor of 2.

\subsection{Limits on the Detectability of \vturb{}}

\FigureMaps
\FigureScaledLinewidths

The single molecule methods, either direct or parametric, are limited by our ability to recover with precision the kinetic temperature. Uncertainty on the kinetic temperature come from different origins: thermal noise, incomplete thermalization of the observed spectral lines, absolute calibration accuracy, and in the parametric model, inadequacy of the model. Thermal noise can be  overcome by sufficient integration time. Incomplete thermalization is a complex issue, and will in general require multi-line transition to be evaluated. However, in the case of CO, the critical densities are low, and we expect the CO lines to be very close to thermalization. Absolute calibration will place an ultimate limit to the our capabilities to measure the turbulence.

We derive in Appendix~\ref{sec:derivation} the impact of the uncertainty on the kinetic temperature on the derivation of the turbulence,

\begin{equation}
\frac{\delta v_{\rm turb}}{v_{\rm turb}} = \frac{\mu_{\rm H}}{2 \mu \mathcal{M}^2} \, \frac{\delta T}{T},
\label{eq:errordv}
\end{equation}

\noindent where \Mach{} is the Mach number of the turbulent broadening. The left panel of Fig.~\ref{fig:limits} shows, in the absence of any error in the measurement of the linewidth, the relative error in \vturb{} as a function of relative error in \Tkin{} for CO (assuming $\mu = 28$). Note that as errors in $\Delta V$ have been neglected, Fig.~\ref{fig:limits}a underestimates the necessary precision in \Tkin{} to detect \vturb{}.

Previous measurements from the Plateau de Bure Interferometer (PdBI) and Sub-Millimetre Array (SMA) have typical flux calibrations of $\sim$~10\% and $\sim$~20\% respectively \citep{Hughes_ea_2011,Guilloteau_ea_2012}, so we estimate that these can only directly detect \vturb{} at $3 \sigma$ when \vturb{} $\gtrsim 0.16~c_s$ and $\gtrsim 0.26~c_s$ respectively. Our current ALMA experiment has 7-10 \% calibration accuracy, thus is sensitive to \vturb{}~$\gtrsim~0.2~c_s$ for the turbulence not to be consistent with \vel{0} to $5~\sigma$. Ultimately, ALMA is expected to reach a flux calibration of $\approx$~3\%, which will translate to a limit of \vturb{} $\gtrsim 0.07~c_s$ for a $\geq 3\sigma$ detection.

However, the flux calibration does not affect the precision to which widths can be measured. The resulting errors on turbulence and temperature derived in the co-spatial method are given by,

\begin{equation}
\frac{\delta v_{\rm turb}}{v_{\rm turb}} = \frac{1}{\mu_{\rm B} - \mu_{\rm A}} \, \frac{\delta \Delta V}{\Delta V} \sqrt{\,\left(\mu_{\rm A} + \frac{\mu_{\rm H}}{\mathcal{M}^2} \right)^2 + \frac{1}{x^2} \, \left( \mu_{\rm B} + \frac{\mu_{\rm H}}{\mathcal{M}^2} \right)^2},
\end{equation}

\begin{equation}
\frac{\delta T}{T} = \frac{2 \mu_{\rm A} \mu_{\rm B}}{\mu_{\rm B} - \mu_{\rm A}}  \frac{\delta \Delta V}{\Delta V} \sqrt{\,\left(  \frac{\mathcal{M}^2}{\mu_{\rm H}} + \frac{1}{\mu_{\rm A}} \right)^2 + \frac{1}{x^2} \, \left( \frac{\mathcal{M}^2}{\mu_{\rm H}} + \frac{1}{\mu_{\rm B}}\right)^2},
\end{equation}

\FigurePrecision

\noindent where $x$ is a scaling factor between the relative errors on the two linewidths,

\begin{equation}
\frac{\delta \Delta V_{\rm A}}{\Delta V_{\rm A}} = x \cdot \frac{\delta \Delta V_{\rm B}}{\Delta V_{\rm B}} = \frac{\delta \Delta V}{\Delta V}. 
\end{equation}

\noindent See the Appendix~\ref{sec:derivation} for the complete derivation. 

Figure \ref{fig:limits}b shows the relative error on \vturb{} assuming the molecular masses of CN and CS (26 and 44 respectively) and that the relative errors on both lines are the same, $x=1$. Figure~\ref{fig:limits}c shows the limits of this method in determining \Tkin{}. For the observations presented in this paper, we have a precision in the measurement of the linewidth of $\approx 0.3~\%$ for both CO and CN, and $\approx 1~\%$ for CS (hence $x \approx 0.33$).

Parametric models typically return much lower formal errors on \vturb{} than direct methods (for example, we find relative errors in Section.~\ref{sec:method3} on the order of 5\%). However, this is only a result of the imposed prior on the shape of the radial dependency of the temperature and turbulent width, which can lead to a significant bias that is not accounted for in the analysis. In any case, these parametric models suffer from the same fundamental limits due to thermalization and absolute calibration as the single-molecule direct method.

\subsection{Comparison with Other Observations, Disks and Simulations}

Turbulence in TW~Hya was modelled previously by \citet{Hughes_ea_2011} using \vel{40} resolution SMA observations of CO (3-2). Using a model fitting approach the authors found an upper limit of \vturb{} $\la$ \vel{40} corresponding to $\la 0.1 c_s$, considerably lower than the values plotted in Fig.~\ref{fig:direct_turbulent}. The temperature profile assumed for their parametric model was warmer than found in this work, with the authors quoting $T_{100} = 40$~K and $e_T = 0.4$ compared to our values of $T_{100} = 34.5 \pm 0.1$~K and $e_T = 0.492 \pm 0.002$ (see Fig.~\ref{fig:temperature_comparisons}). This warmer profile is sufficient to account for any difference in resulting \vturb{}. In any case, both measurements are fundamentally limited by the absolute calibration uncertainty, and only imply \vturb{}~$< 0.23 c_s$ (SMA data) or $< 0.16 c_s$ (our ALMA data).

Other disks have also been the subject of investigations of \vturb{}. DM~Tau, MWC~480 and LkCa~15 have yielded higher velocities of $\la 100 - 200$~m\,s$^{-1}$ ($\la 0.3-0.5$~$c_s$) \citep{Dartois_ea_2003,Pietu_ea_2007} which are sufficiently large to be detected by the PdBI. However, the velocity resolution of the observations was on the order of \vel{200} resulting in a poorly constrained total linewidth which may result in overestimating \vturb{}. The impact of the spectral resolution was accounted for in the more recent measurement of DM Tau by \citet{Guilloteau_ea_2012} using the heavier molecule CS, who found \vturb{}~$\simeq 0.3 - 0.4~c_s$ More recently \citet{Flaherty_ea_2015} used parametric modeling of multiple CO isotopologue transitions to infer \vturb{} $\la 0.04~c_s$ in HD~163296. 

We must, however, consider the impact of flux calibration on all methods involving a single line measurement. Every method will constrain the local linewidth using some combination of diagnostics, such as the broadening of channel images or the peak-to-trough ratio of the integrated spectra. Each method will recover this linewidth to its own precision (depending particularly on the functional form imposed for the spatial dependency of this linewidth). However, once the uncertainty on the local linewidth is known, Equation~\ref{eq:errordv} can be applied to propagate the error due to this uncertainty and to the absolute calibration precision to the turbulent component of the line width. Application to the results of \citet{Hughes_ea_2011} and \citet{Flaherty_ea_2015} yield upper limits of \vturb{}~$< 0.23~c_s$ and $< 0.16~c_s$ respectively, more similar to what we measure here. The \vturb{} value found for DM~Tau is considerably larger than limits imposed by the flux calibration $(\approx 10\%$). Given consideration of the observed limits, this suggests that the disk of DM~Tau is more turbulent than those of TW Hya and HD~163296.

Comparisons with numerical simulations also provide a chance to distinguish between turbulent mechanisms. \citet{Simon_ea_2015} used an ensemble of shearing-box MHD simulations coupled with radiative transfer modeling to predict the velocity dispersion traced by CO emission in a proto-typical T-Tauri disk pervaded by MRI. The authors found that molecular emission would trace a transition region between the dead-zone and the turbulent atmosphere, showing velocity dispersion of between 0.1 and 0.3~$c_s$, almost identical to the range found in TW~Hya. \citet{Flock_ea_2015} ran similar, however global, models of an MRI active disk, finding velocity dispersions of \vturb{} $\approx$ 40 -- \vel{60} near the midplane, rising to 80 -- \vel{120} higher above the midplane, again consistent with the values found in TW~Hya. A comparison with the $\alpha$ viscosity models is more complex, as the relation between \vturb{} and $\alpha$ depends on the nature of the viscosity, with \vturb{} ranging between a few $\alpha c_s$ and $\sqrt{\alpha} c_s$ \citep{Cuzzi_ea_2001}.

A vertical dependence of \vturb{}, as found in \citet{Flock_ea_2015}, is a typical feature of MRI driven turbulence and may provide a discriminant between other models of turbulent mixing. In addition to the parametric model finding different temperatures for all three molecules, CO and CN yielded different \Tex{} values from the line profile fitting and the simultaneous method failed under the assumption that CN and CS are co-spatial. These pieces of evidence suggest CO, CN and CS each trace distinct vertical regions in the disk, potentially providing a possibility to trace a vertical gradient in \vturb{}. With the current uncertainties on the temperatures for the three molecules we are unable to distinguish any difference in \vturb{} with height above the midplane.

\citet{Cleeves_ea_2015} have modelled the ionization structure of TW~Hya using observations of key molecular ions HCO$^+$ and N$_2$H$^+$, concluding that the disk may have a large MRI-dead zone extending to $\sim 50-65$~au. An observable feature of such a dead zone would be a sharp decrease in the velocity dispersion at this radius. Our data lack the spatial resolution and sensitivity to reliably trace the gas turbulent motions in the inner $\sim 40$~au where this feature may be more prominent. However, the power law analysis indicates \vturb values actually increase with radius (exponent $e_{\delta\mathrm{v}} < 0$), in contrast with the direct measurements. This difference may be due to the impact of such a less turbulent inner region that is ignored in the direct method, but must be fitted in the power law analysis.

Future observations will improve the above analysis: in order to improve the accuracy of \vturb{} determination with this direct method, a well constrained thermal structure is crucial. This can be attained with observations of multiple transitions of the same molecule. Furthermore, for more inclined systems, a better understanding of the impact of beam smearing on the velocity dispersion is paramount. This can be combated with smaller beamsizes, resolving a smaller shear component. Among observed species, CS currently provides the best opportunity to probe velocity dispersions closer to the midplane, while we have demonstrated that the ensemble of CO, CN and CS can allow additionally for the determination of the vertical dependence of \vturb{}. Despite all these improvements, direct measures of turbulence will ultimately be limited by the flux calibration of the interferometers with a sensitivity of $\approx 0.1 c_s$ for ALMA's quoted 3\% accuracy.

\section{Conclusion}
\label{sec:conclusion}

\FigureTemperatureComparison

We have discussed several methods of obtaining the turbulent velocity dispersion in the disk of TW~Hya using CO, CN and CS rotational emission with a view to complementing the commonly used parametric modeling approach. Guided by previous models of TW~Hya, the direct method yields \vturb{} values which depend strongly on the radius of the disk, reaching $\approx$~\vel{150} at 40~au, dropping to a near constant $\approx$~\vel{50} outside 100~au for all three tracers. As a function of local soundspeed, CO and CN displayed a near constant \vturb{} $\sim 0.2~c_s$. However, the analysis of the possible sources of errors shows that these numbers should most likely be interpreted as upper limits.

Direct or parametric methods using a single molecule are limited by a poor knowledge of the thermal structure of the disk. Additional transition lines will provide a more accurate determination of the temperature, however this is ultimately limited by the flux calibration of ALMA. With an expected minimum of 3\% error on flux calibration, we estimate that a firm detection of turbulent broadening is only possible if \vturb{} $/ c_s \ga 0.1$ via this direct method. The co-spatial method can potentially overcome this absolute calibration issue, however requires two co-spatial tracers of sufficient abundance to have strong emission. Tracing \vturb{} close to the midplane will be considerably more challenging, requiring a strong detection of o-H$_2$D$^+$ and another molecule residing in the midplane, such as N$_2$D$^+$.

\begin{acknowledgements}
We thank the referee who's helpful comments have improved this manuscript. R.T. is a member of the International Max Planck Research School for Astronomy and Cosmic Physics at the University of Heidelberg, Germany. D. S. and T. B. acknowledge support by the Deutsche Forschungsgemeinschaft through SPP 1385: "The first ten million years of the solar system a planetary materials approach" (SE 1962/1- 3) and SPP 1833 "Building a Habitable Earth (KL 1469/13-1), respectively. This research made use of  System. This paper makes use of the following ALMA data: ADS/JAO.ALMA\#2013.1.00387.S. ALMA is a partnership of ESO (representing its member states), NSF (USA) and NINS (Japan), together with NRC (Canada), NSC and ASIAA (Taiwan), and KASI (Republic of Korea), in cooperation with the Republic of Chile. The Joint ALMA Observatory is operated by ESO, AUI/NRAO and NAOJ. This work was supported by the National Programs PCMI and PNPS from INSU-CNRS.
\end{acknowledgements}

\bibliographystyle{aa}
\bibliography{main}

\begin{thebibliography}{24}
\expandafter\ifx\csname natexlab\endcsname\relax\def\natexlab#1{#1}\fi

\bibitem[{{Brinch} \& {Hogerheijde}(2010)}]{Brinch_ea_2010}
{Brinch}, C. \& {Hogerheijde}, M.~R. 2010, \aap, 523, A25

\bibitem[{{Cleeves} {et~al.}(2015){Cleeves}, {Bergin}, \&
  {Harries}}]{Cleeves_ea_2015}
{Cleeves}, L.~I., {Bergin}, E.~A., \& {Harries}, T.~J. 2015, \apj, 807, 2

\bibitem[{{Cuzzi} {et~al.}(2001){Cuzzi}, {Hogan}, {Paque}, \&
  {Dobrovolskis}}]{Cuzzi_ea_2001}
{Cuzzi}, J.~N., {Hogan}, R.~C., {Paque}, J.~M., \& {Dobrovolskis}, A.~R. 2001,
  \apj, 546, 496

\bibitem[{{Dartois} {et~al.}(2003){Dartois}, {Dutrey}, \&
  {Guilloteau}}]{Dartois_ea_2003}
{Dartois}, E., {Dutrey}, A., \& {Guilloteau}, S. 2003, \aap, 399, 773

\bibitem[{{Dutrey} {et~al.}(2011){Dutrey}, {Wakelam}, {Boehler}, {Guilloteau},
  {Hersant}, {Semenov}, {Chapillon}, {Henning}, {Pi{\'e}tu}, {Launhardt},
  {Gueth}, \& {Schreyer}}]{Dutrey_ea_2011}
{Dutrey}, A., {Wakelam}, V., {Boehler}, Y., {et~al.} 2011, \aap, 535, A104

\bibitem[{{Flaherty} {et~al.}(2015){Flaherty}, {Hughes}, {Rosenfeld},
  {Andrews}, {Chiang}, {Simon}, {Kerzner}, \& {Wilner}}]{Flaherty_ea_2015}
{Flaherty}, K.~M., {Hughes}, A.~M., {Rosenfeld}, K.~A., {et~al.} 2015, \apj,
  813, 99

\bibitem[{{Flock} {et~al.}(2015){Flock}, {Ruge}, {Dzyurkevich}, {Henning},
  {Klahr}, \& {Wolf}}]{Flock_ea_2015}
{Flock}, M., {Ruge}, J.~P., {Dzyurkevich}, N., {et~al.} 2015, \aap, 574, A68

\bibitem[{{Gorti} {et~al.}(2011){Gorti}, {Hollenbach}, {Najita}, \&
  {Pascucci}}]{Gorti_ea_2011}
{Gorti}, U., {Hollenbach}, D., {Najita}, J., \& {Pascucci}, I. 2011, \apj, 735,
  90

\bibitem[{{Guilloteau} {et~al.}(2012){Guilloteau}, {Dutrey}, {Wakelam},
  {Hersant}, {Semenov}, {Chapillon}, {Henning}, \&
  {Pi{\'e}tu}}]{Guilloteau_ea_2012}
{Guilloteau}, S., {Dutrey}, A., {Wakelam}, V., {et~al.} 2012, \aap, 548, A70

\bibitem[{{Henning} \& {Meeus}(2011)}]{Henning_Meeus_2011}
{Henning}, T. \& {Meeus}, G. 2011, {Dust Processing and Mineralogy in
  Protoplanetary Accretion Disks}, ed. P.~J.~V. {Garcia} (Chicago: University
  of Chicago Press), 114--148

\bibitem[{{Hughes} {et~al.}(2011){Hughes}, {Wilner}, {Andrews}, {Qi}, \&
  {Hogerheijde}}]{Hughes_ea_2011}
{Hughes}, A.~M., {Wilner}, D.~J., {Andrews}, S.~M., {Qi}, C., \& {Hogerheijde},
  M.~R. 2011, \apj, 727, 85

\bibitem[{{Lenz} \& {Ayres}(1992)}]{Lenz+Ayres_1992}
{Lenz}, D.~D. \& {Ayres}, T.~R. 1992, \pasp, 104, 1104

\bibitem[{{M{\"u}ller} {et~al.}(2001){M{\"u}ller}, {Thorwirth}, {Roth}, \&
  {Winnewisser}}]{CDMS_2001}
{M{\"u}ller}, H.~S.~P., {Thorwirth}, S., {Roth}, D.~A., \& {Winnewisser}, G.
  2001, \aap, 370, L49

\bibitem[{{Pi{\'e}tu} {et~al.}(2007){Pi{\'e}tu}, {Dutrey}, \&
  {Guilloteau}}]{Pietu_ea_2007}
{Pi{\'e}tu}, V., {Dutrey}, A., \& {Guilloteau}, S. 2007, \aap, 467, 163

\bibitem[{{Pringle}(1981)}]{Pringle81}
{Pringle}, J.~E. 1981, \araa, 19, 137

\bibitem[{{Qi} {et~al.}(2004){Qi}, {Ho}, {Wilner}, {Takakuwa}, {Hirano},
  {Ohashi}, {Bourke}, {Zhang}, {Blake}, {Hogerheijde}, {Saito}, {Choi}, \&
  {Yang}}]{Qi_ea_2004}
{Qi}, C., {Ho}, P.~T.~P., {Wilner}, D.~J., {et~al.} 2004, \apjl, 616, L11

\bibitem[{{Qi} {et~al.}(2013){Qi}, {{\"O}berg}, {Wilner}, {D'Alessio},
  {Bergin}, {Andrews}, {Blake}, {Hogerheijde}, \& {van Dishoeck}}]{Qi_ea_2013}
{Qi}, C., {{\"O}berg}, K.~I., {Wilner}, D.~J., {et~al.} 2013, Science, 341, 630

\bibitem[{{Roberge} {et~al.}(2005){Roberge}, {Weinberger}, \&
  {Malumuth}}]{Roberge_ea_2005}
{Roberge}, A., {Weinberger}, A.~J., \& {Malumuth}, E.~M. 2005, \apj, 622, 1171

\bibitem[{{Rosenfeld} {et~al.}(2013){Rosenfeld}, {Andrews}, {Hughes}, {Wilner},
  \& {Qi}}]{Rosenfeld_ea_2013}
{Rosenfeld}, K.~A., {Andrews}, S.~M., {Hughes}, A.~M., {Wilner}, D.~J., \&
  {Qi}, C. 2013, \apj, 774, 16

\bibitem[{{Rosenfeld} {et~al.}(2012){Rosenfeld}, {Qi}, {Andrews}, {Wilner},
  {Corder}, {Dullemond}, {Lin}, {Hughes}, {D'Alessio}, \&
  {Ho}}]{Rosenfeld_ea_2012}
{Rosenfeld}, K.~A., {Qi}, C., {Andrews}, S.~M., {et~al.} 2012, \apj, 757, 129

\bibitem[{{Shakura} \& {Sunyaev}(1973)}]{Shakura_ea_1973}
{Shakura}, N.~I. \& {Sunyaev}, R.~A. 1973, \aap, 24, 337

\bibitem[{{Simon} {et~al.}(2015){Simon}, {Hughes}, {Flaherty}, {Bai}, \&
  {Armitage}}]{Simon_ea_2015}
{Simon}, J.~B., {Hughes}, A.~M., {Flaherty}, K.~M., {Bai}, X.-N., \&
  {Armitage}, P.~J. 2015, \apj, 808, 180

\bibitem[{{Skatrud} {et~al.}(1983){Skatrud}, {De Lucia}, {Blake}, \&
  {Sastry}}]{Skatrud_ea_1983}
{Skatrud}, D.~D., {De Lucia}, F.~C., {Blake}, G.~A., \& {Sastry}, K.~V.~L.~N.
  1983, Journal of Molecular Spectroscopy, 99, 35

\bibitem[{{Testi} {et~al.}(2014){Testi}, {Birnstiel}, {Ricci}, {Andrews},
  {Blum}, {Carpenter}, {Dominik}, {Isella}, {Natta}, {Williams}, \&
  {Wilner}}]{Testi_ea_2014}
{Testi}, L., {Birnstiel}, T., {Ricci}, L., {et~al.} 2014, {Dust Evolution in
  Protoplanetary Disks}, ed. H.~{Beuther}, C.~{Dullemond}, R.~{Klessen}, \&
  T.~{Henning}, 339--361

\end{thebibliography}

\newpage
\begin{appendix}

\section{CN Hyperfine Components}
\label{sec:cnhyperfine}

\CNVelocitiesTable

Here we present the relative offsets of the CN N=(2-1) hyperfine components used in the paper. The old and new values are given in Table.~\ref{tab:cn21}.

\section{Error Derivations}
\label{sec:derivation}

In this section we discuss the uncertainties arising from the line profile fitting and their subsequent propagation into the derivation of \vturb{}.

\subsection{Uncertainties from Line Profile Fitting}

\begin{figure}
\centering
\includegraphics[width=0.45\textwidth]{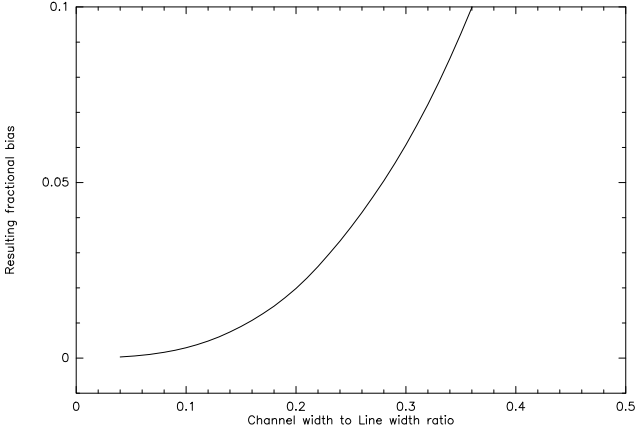}
\caption{The impact of spectrally resolving a line on the determination of $\Delta V$. For the data presented in this paper, CO and CN have a maximum ratio of 0.21 and CS has 0.25 resulting in fractional errors for $\Delta V$ of $\approx 2~\%$ and $3.5~\%$ respectively.}
\label{fig:resolution}
\end{figure}

The uncertainty of a Gaussian line parameter $X$ is derived from Eq.~1 from \citet{Lenz+Ayres_1992},

\begin{equation}
\frac{\delta X}{X} = \frac{1}{C_X} \sqrt{\frac{\delta v}{\Delta V}} \frac{\sigma}{T_p}
\end{equation}

\noindent where $C_X$ is a coefficient of order 1, given in Table 1 by the same authors. $\sigma$ is the noise per channel of width $\delta v$, and $T_p$ the peak intensity of the Gaussian, while $\Delta V$ is its FWHM.  For the line width, $C_X \approx 0.6$. The simultaneous fit of the opacity slightly reduces this number.

In our observations, in one channel of 30.5~kHz, or about \vel{40}, we have a typical rms noise of order 6 - 7 mJy/beam which translates to about 0.5 - 0.7 K at the angular resolution of our data (around $0.5\arcsec$).  CO has line widths around \vel{300}, and a peak intensity around 40~K, yielding errors on the line widths about \vel{2.7} (1~\% precision) for each beam. The CN peak typical brightness is lower, about 20~K, but CN has several hyperfine components, so that the precision obtained from CN is only 1.5 times worse than from CO. For CS, the peak brightness is rather on the order of 10~K, leading to a precision of about 4\%.

The error is further reduced by azimuthal averaging, since the number of independent beams increases as $\sqrt(r)$. At 100~au, we average about 22 beams, and the gain is about a factor 4.7. Thus the final precision on the linewidths on the azimuthal average is respectively of the order 1~\% for CS, and a factor 2 (resp. 3) better for CN (resp. CO).

It is also worth noting that for a given integration time, the precision does not depend on the selected spectral resolution, provided it is sufficient to sample the lineshape. Fig.~\ref{fig:resolution} demonstrates the incurred bias when the line is not sufficiently resolved.

\subsection{Direct Turbulent Velocity Dispersion}

Assuming \Tkin{} is known, then the turbulent velocity component $v_{\rm turb}$ is given by:

\begin{equation}
v_{\rm turb} = \sqrt{\Delta V^2 - \frac{2kT}{\mu m_{\rm p}}}.
\label{eq:vturb}
\end{equation}

\noindent We make the assumption that $\delta T \gg \delta \Delta V$, so the uncertainty on $v_{\rm turb}$ is:

\begin{align}
\delta v_{\rm turb} &= \left| \,\frac{\partial v_{\rm turb}}{\partial T} \delta T \,\right|, \\
&= \frac{k}{\mu m_{\rm p}} \, \left( \Delta V^2 - \frac{2kT}{\mu m_{\rm p}} \right)^{-1/2} \delta T.
\end{align}

\noindent Dividing through by $v_{\rm turb}$ in order to obtain the relative error gives:

\begin{equation}
\frac{\delta v_{\rm turb}}{v_{\rm turb}} = \frac{k}{\mu m_{\rm p}} \, \left( \Delta V^2 - \frac{2kT}{\mu m_{\rm p}} \right)^{-1} \delta T.
\label{eq:dvv_a}
\end{equation}

\noindent From rearranging Equation~\ref{eq:vturb} for $\Delta V$,

\begin{align}
\Delta V_i^2 &= v_{\rm turb}^2 + \frac{2kT}{\mu_i m_{\rm p}} \\
 &= \frac{2kT}{m_{\rm p}} \left( \frac{\mathcal{M}^2}{\mu_{\rm H}} + \frac{1}{\mu_i}\right) \quad \text{where} \quad \mathcal{M} \equiv \frac{v_{\rm turb}}{\sqrt{2} c_s},
\label{eq:totalwidth}
\end{align}

\noindent we can be substitute this into Equation~\ref{eq:dvv_a} to yield:

\begin{align}
\frac{\delta v_{\rm turb}}{v_{\rm turb}} &= \frac{k}{\mu m_{\rm p}} \, \left( \frac{2kT}{m_{\rm p}} \left[ \frac{\mathcal{M}^2}{\mu_{\rm H}} + \frac{1}{\mu}\right] - \frac{2kT}{\mu m_{\rm p}} \right)^{-1} \delta T, \\
&= \frac{2\mu_{\rm H}}{\mu \mathcal{M}^2} \, \frac{\delta T}{T}.
\end{align}

\subsection{Co-Spatial Kinetic Temperature}

The kinetic temperature and its associated uncertainty are:

\begin{align}
T &= \frac{m_{\rm p}}{2k} \frac{\mu_{\rm a} \mu_{\rm b}}{\mu_{\rm b} - \mu_{\rm a}}  \left( \Delta V^2_{\rm a} -  \Delta V^2_{\rm b} \right), \label{eq:t}\\
\delta T &= \frac{2 m_{\rm p}}{2k} \frac{\mu_{\rm A} \mu_{\rm B}}{\mu_{\rm B} - \mu_{\rm A}} \sqrt{\Big( \Delta V_{\rm A} \cdot \delta \Delta V_{\rm A}\Big)^2 + \Big( \Delta V_{\rm B} \cdot \delta \Delta V_{\rm B}\Big)^2}. \label{eq:dt}
\end{align}

\noindent We assume that the relative errors on the linewidth are proportional to one another such that,

\begin{equation}
\frac{\delta \Delta V_{\rm A}}{\Delta V_{\rm A}} = x \cdot \frac{\delta \Delta V_{\rm B}}{\Delta V_{\rm B}} = \frac{\delta \Delta V}{\Delta V}, 
\label{eq:relwidths}
\end{equation}

\noindent where $x$ scales the relative errors if they are not the same; Fig.~\ref{fig:limits} uses $x=1$. Substituting these into the Equation~\ref{eq:dt}:

\begin{align}
\delta T &= 2\,\frac{m_{\rm p}}{2k} \frac{\mu_{\rm A} \mu_{\rm B}}{\mu_{\rm B} - \mu_{\rm A}} \sqrt{\,\left( \Delta V_{\rm A}^2 \cdot \frac{\delta \Delta V}{\Delta V}\right)^2 + \, \left( \frac{\Delta V_{\rm B}^2}{x} \cdot \frac{\delta \Delta V}{\Delta V}\right)^2}, \\
&= 2\,\frac{m_{\rm p}}{2k} \frac{\mu_{\rm A} \mu_{\rm B}}{\mu_{\rm B} - \mu_{\rm A}} \frac{\delta \Delta V}{\Delta V} \sqrt{\Delta V_{\rm A}^4 + \frac{\Delta V_{\rm B}^4}{x^2}}.
\end{align}

\noindent Substitute in for $\Delta V_{\rm A}^4$ and $\Delta V_{\rm B}^4$ from Equation~\ref{eq:totalwidth} and rearrange for the relative uncertainty on $T$:

\begin{equation}
\frac{\delta T}{T} = \frac{2 \mu_{\rm A} \mu_{\rm B}}{\mu_{\rm B} - \mu_{\rm A}}  \frac{\delta \Delta V}{\Delta V} \sqrt{\left(  \frac{\mathcal{M}^2}{\mu_{\rm H}} + \frac{1}{\mu_{\rm A}}\right)^2 + \frac{1}{x^2} \cdot \left( \frac{\mathcal{M}^2}{\mu_{\rm H}} + \frac{1}{\mu_{\rm B}}\right)^2}.
\end{equation}

\subsection{Co-Spatial Turbulent Velocity Dispersion}

\noindent We can play the same game with the turbulent velocity dispersion:

\begin{align}
v_{\rm turb} &= \sqrt{\frac{\mu_{\rm B} \Delta V^2_{\rm B} - \mu_{\rm A} \Delta V^2_{\rm A}} { \mu_{\rm B} - \mu_{\rm A}}} \label{eq:vturb_a}\\
\delta v_{\rm turb} &= \sqrt{ \frac{ (\mu_{\rm B} \Delta V_{\rm B} \delta \Delta V_{\rm B})^2 + (\mu_{\rm A} \Delta V_{\rm A} \delta \Delta V_{\rm A})^2} { (\mu_{\rm B} - \mu_{\rm A}) \cdot (\mu_{\rm B} \Delta V_{\rm B}^2 - \mu_{\rm A} \Delta V_{\rm A}^2)}}.
\end{align}

\noindent Substitute in for the relative linewidths from Equation~\ref{eq:relwidths} and for $v_{\rm turb}$ from Equation~\ref{eq:vturb} to give:

\begin{equation}
\delta v_{\rm turb} = \frac{1}{\mu_{\rm B} - \mu_{\rm A}} \, \frac{1}{v_{\rm turb}} \, \frac{\delta \Delta V}{\Delta V} \sqrt{\mu_{\rm A}^2 \Delta V^4_{\rm A} + \frac{\mu_{\rm B}^2}{x^2} \Delta V^4_{\rm B}}.
\label{eq:vturb_b}
\end{equation}

\noindent Rearranging Equation~\ref{eq:totalwidth} yields:

\begin{equation}
\mu_i \Delta V_i^2 = v_{\rm turb}^2 \left( \mu_i + \frac{\mu_{\rm H}}{\mathcal{M}^2} \right),
\end{equation}

\noindent which can be substituted into Equation~\ref{eq:vturb_b}. After some rearranging we find:

\begin{equation}
\frac{\delta v_{\rm turb}}{v_{\rm turb}} = \frac{1}{\mu_{\rm B} - \mu_{\rm A}} \, \frac{\delta \Delta V}{\Delta V} \sqrt{\left( \mu_{\rm A} + \frac{\mu_{\rm H}}{\mathcal{M}^2} \right)^2 + \frac{1}{x^2} \cdot \left( \mu_{\rm B} + \frac{\mu_{\rm H}}{\mathcal{M}^2} \right)^2}.
\end{equation}

\section{Observations}
\label{sec:appobservations}

\FigureCNCSMaps

\begin{figure*}
\centering
\includegraphics[width=0.9\textwidth]{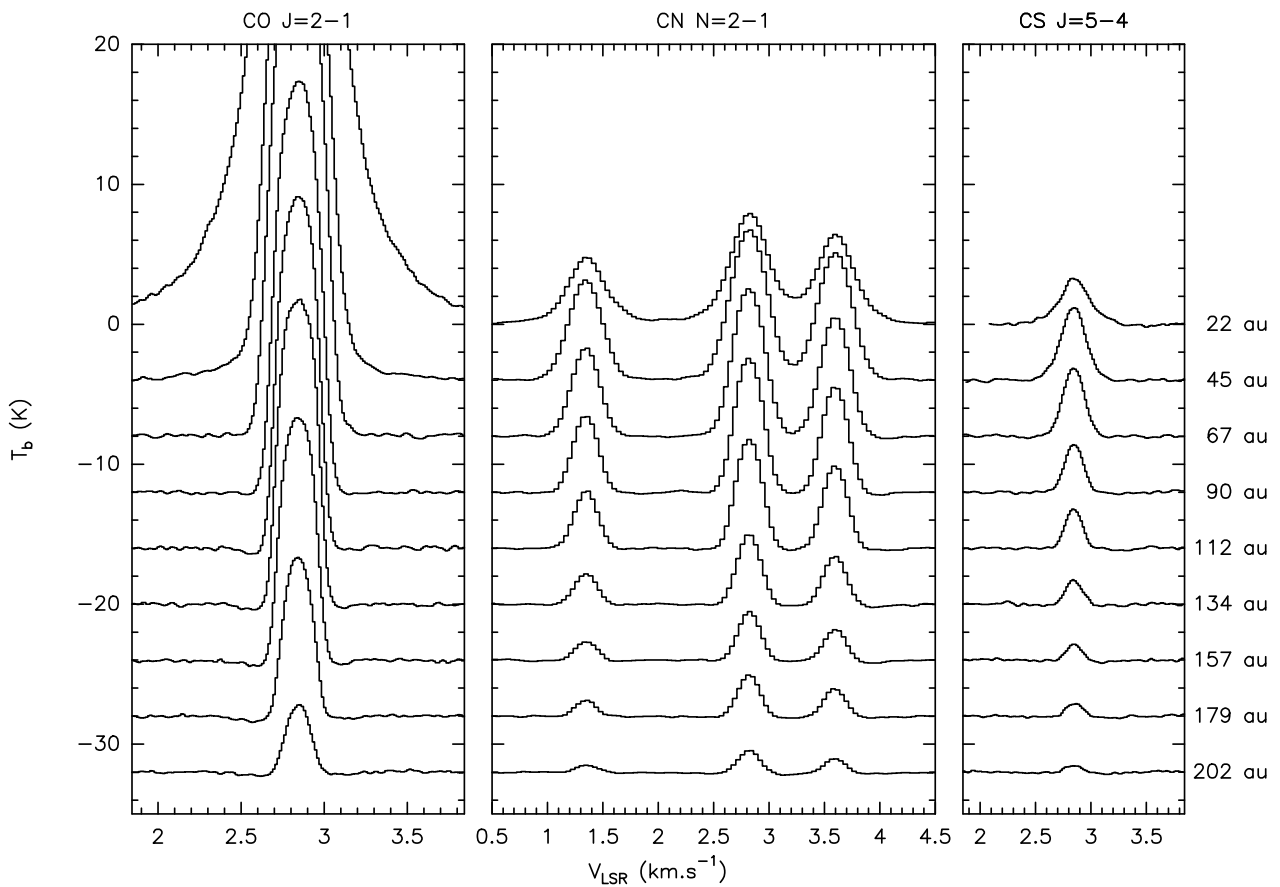}
\caption{Showing the azimuthally averaged spectra for the three emission lines: CO, left; CN, centre; and CS, right. The radial sampling is roughly a beamsize.}
\label{fig:aaspectra}
\end{figure*}

\begin{figure*}
\centering
\includegraphics[width=0.9\textwidth]{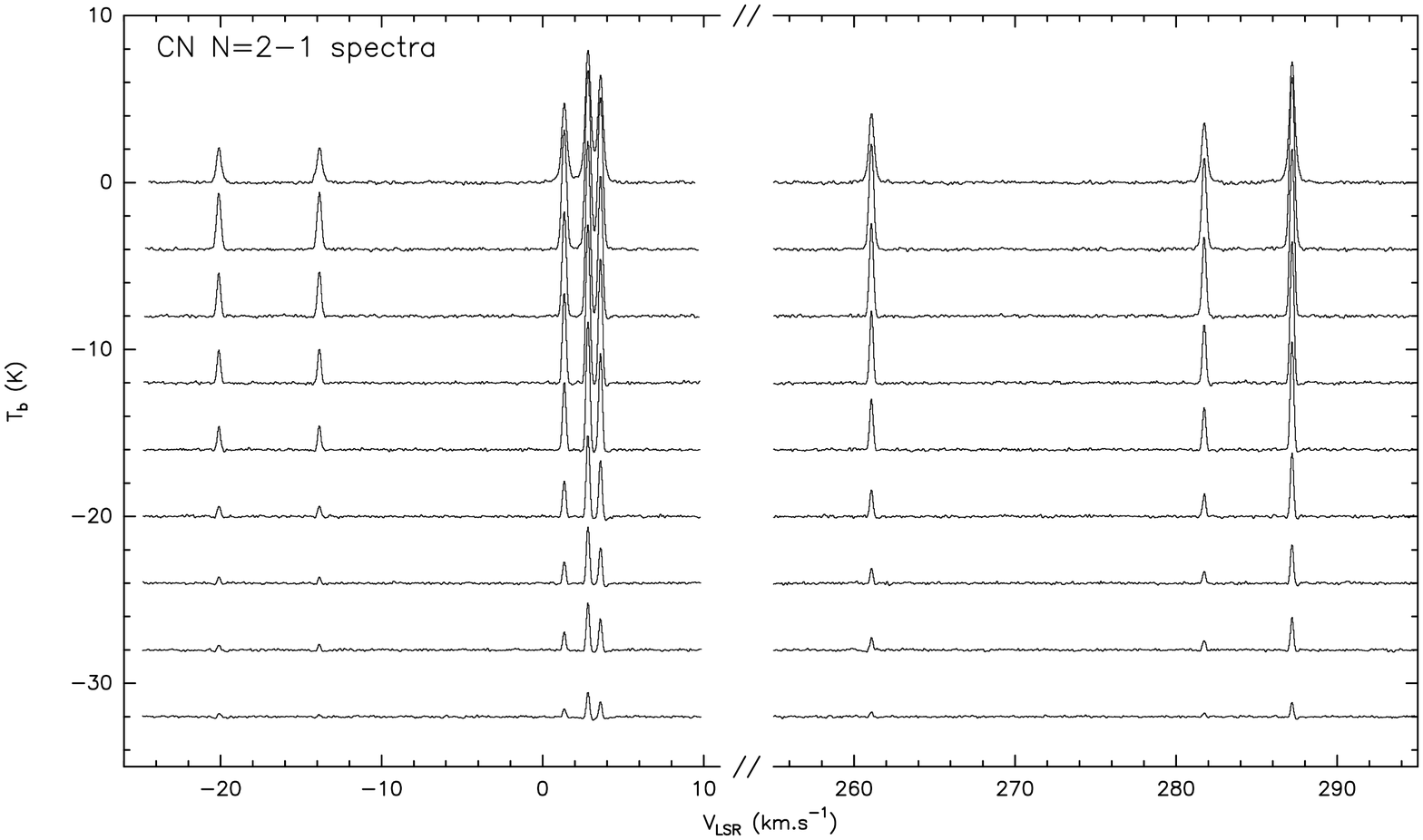}
\caption{The full compliment of CN = (2-1) hyperfine components as described in Table~\ref{tab:cn21}. Each line is an azimuthal average taken at a radius as shown in Fig.~\ref{fig:aaspectra}, roughly a beamsize in distance.}
\label{fig:aaallspectra}
\end{figure*}

Demonstration of the observational data used. Figure~\ref{fig:CNCS_integratedintensity} shows the integrated intensities of CN (including all hyperfine components) and CS clearly demonstrating the lack of azimuthal structure, as with CO which is identical to previous studies. 

Examples of the spectra used for the analysis are shown in Fig.~\ref{fig:aaspectra}. The full compliment of hyperfine components can be found in Fig.~\ref{fig:aaallspectra}.

\end{appendix}
\end{document}